\documentclass[12pt]{article}

\usepackage[utf8]{inputenc}
\usepackage{amsmath,mathtools,amsfonts}
\usepackage{bm}
\usepackage[dvipdfmx]{graphicx}
\usepackage[dvipdfmx]{graphics}
\usepackage[all]{xy}
\usepackage{mathrsfs}
\usepackage{cite}

\newcommand{\sectiono}[1]{\section{#1}\setcounter{equation}{0}}

\newcommand{\bd}[1]{\boldsymbol{#1}}
\newcommand{\id}{\mathbb{I}}

\newcommand{\NSNS}{NS\textrm{-}NS}
\newcommand{\RNS}{R\textrm{-}NS}
\newcommand{\NSR}{NS\textrm{-}R}
\newcommand{\RR}{R\textrm{-}R}
\newcommand{\sst}{(s,\bar{s},t)}
\newcommand{\ssb}{(s,\bar{s})}

\begin{document}

\baselineskip=17pt

\begin{titlepage}
\rightline{\tt YITP-22-32}
\begin{center}
\vskip 2.5cm
{\Large \bf {Open-closed homotopy algebra\\ 
in superstring field theory}}
\vskip 1.0cm
{\large {Hiroshi Kunitomo}}
\vskip 1.0cm
{\it {Center for Gravitational Physics and Quantum Information}},\\ 
{\it {Yukawa Institute for Theoretical Physics}}
{\it {Kyoto University}},\\
{\it {Kitashirakawa Oiwakecho, Sakyo-ku, Kyoto 606-8502, Japan}}\\
kunitomo@yukawa.kyoto-u.ac.jp

\vskip 2.0cm

{\bf Abstract}
\end{center}
\noindent

We construct open-closed superstring interactions based on the open-closed
homotopy algebra structure. It provides a classical open superstring field theory 
on general closed-superstring-field backgrounds described by classical solutions 
of the nonlinear equation of motion of the closed superstring field theory.
We also give the corresponding WZW-like action through the map connecting
the homotopy-based and WZW-like formulations.

\end{titlepage}

\tableofcontents

\newpage

\sectiono{Introduction}

It is known that several homotopy algebras are naturally realized as algebraic structures 
in string field theories and play a significant role. This was first recognized in closed 
bosonic string field theory \cite{Zwiebach:1992ie,Lada:1992wc}, 
where the $L_\infty$ structure determines the (classical) gauge-invariant action.
Open bosonic string field theory was first formulated as a cubic theory using the 
(Witten) associative product \cite{Witten:1985cc} but can be extended to that 
with an $A_\infty$ structure more generally \cite{Nakatsu:2001da,Kajiura:2003ax}.
This is also deformed to the theory on general closed string backgrounds 
\cite{Zwiebach:1990qj,Zwiebach:1997fe,Zwiebach:1992bw} based on
the open-closed homotopy algebra (OCHA) structure 
\cite{Kajiura:2004xu,Kajiura:2005sn,ocha coalgebra}.
In the superstring field theories, the homotopy algebra structure is more important. 
Since it seems inevitable to avoid associativity anomaly \cite{Wendt:1987zh}, the $A_\infty$
structure becomes essential to determine the gauge-invariant action in the open superstring 
field theory \cite{Erler:2013xta,Erler:2016ybs,Kunitomo:2020xrl}. 
The $L_\infty$ structure again plays the role of guiding principle to determine the action 
with appropriate picture numbers in the heterotic and type II superstring field theories
\cite{Erler:2014eba,Jurco:2013qra,Kunitomo:2019glq,Kunitomo:2019kwk,Kunitomo:2021wiz}.


On the other hand, the current understanding is that there is no essential difference 
between the theory of open string/closed string mixed system and the theory of purely 
closed string. It merely describes the perturbation on the different backgrounds, 
those with and without a D-brane \cite{Polchinski:1995mt,Polchinski:1996fm}.
They should be derived from non-perturbatively formulated fundamental theory such as 
string field theory, but it is not a priori clear which one should be considered more 
fundamental. The closed string field theory is simpler, but the open-closed string 
field theory has a larger symmetry structure, the OCHA structure\footnote{
In a formulation that introduces an auxiliary degree of freedom,
the open-closed superstring field theory has already been constructed 
\cite{FarooghMoosavian:2019yke}.
It does not, however, decrease the worthwhile to construct the theory based on
the OCHA structure.}. The purpose of 
this paper is to construct an open-closed string field theory realizing the OCHA 
structure. The action obtained explains the classical open string field theory 
on general closed-string backgrounds.

The paper is organized as follows. In section \ref{A}, we briefly review 
the open superstring field theory with general $A_\infty$ structure.
After introducing some conventions and fundamental ingredients, we show how 
we construct the open superstring field theory based on the $A_\infty$ 
structure. The superstring products with appropriate picture numbers 
satisfying the $A_\infty$ relations can be 
obtained by recursively solving the differential equations.
We similarly review the closed superstring field theory with the $L_\infty$ structure 
in section \ref{B}.
We define the string products multiplying both open and closed string field
in section \ref{C} and show the relations they must satisfy to form
the OCHA. We also give the differential equations that the products
with OCHA structure should follow. They provide an action
of the open superstring field theory on the general closed superstring backgrounds.
In section \ref{D}, we obtain, as a byproduct, the corresponding WZW-like action through
the map connecting the homotopy-based and WZW-like formulations,
which is a generalization considered in \cite{Maccaferri:2021lau}.
Section \ref{E} is devoted to the summary and
discussion. 
In order to help the construction of concrete string interactions,
some explicit procedure for solving the differential equations is given 
in Appendix \ref{App A0}. 
Appendix \ref{App A} is added to make the paper self-contained.
We introduce two composite string fields, the pure-gauge open string field
and the associated open string field, which is nontrivial in the theory with 
general $A_\infty$ structure.

\sectiono{Open superstring field theory with $A_\infty$-structure}\label{A}

We summarize in this section how the open superstring field theory is 
constructed based on the $A_\infty$ algebra structure.

\subsection{Open superstring field}

The first-quantized Hilbert space 
of open superstring is composed of two sectors: 
$\mathcal{H}_o = \mathcal{H}_{NS} + \mathcal{H}_R$.
Correspondingly, 
the open superstring field $\Psi$ has two components: $\Psi = \Psi_{NS} + \Psi_R$,
both of which are Grassmann odd and have ghost number $1$.
The component $\Psi_{NS}$ ($\Psi_R$) has picture number $-1$ ($-1/2$)
and represents space-time bosons (fermions).
We impose on it a constraint 
\begin{equation}
\mathcal{P}^o_{XY}\Psi\ =\ \Psi,\qquad
\mathcal{P}^o_{XY}\ =\ \mathcal{G}^o(\mathcal{G}^o)^{-1},
\label{open constraint}
\end{equation}
with
\begin{equation}
 \mathcal{G}^o\ =\ \pi^0 + X^o\pi^1,\qquad
 (\mathcal{G}^o)^{-1}\ =\ \pi^0 + Y^o\pi^1,
\end{equation}
where  $\pi^0$ and $\pi^1$ are the projection operators onto the NS and R components,
respectively: $\pi^0\Psi = \Psi_{NS}$ and $\pi^1\Psi = \Psi_R$. The picture changing operator
(PCO) of open superstring $X^o$ and its inverse $Y^o$ are defined by
\begin{equation}
 X^o\ =\ -\delta(\beta_0)G + (\gamma_0\delta(\beta_0)+\delta(\beta_0)\gamma_0)b_0,\qquad 
 Y^o\ =\ - \frac{G}{L_0}\delta(\gamma_0).
\end{equation}
The PCO $X^o$ is BRST exact in the large Hilbert space:
\begin{equation}
X^o=[Q,\Xi^o],\qquad \Xi^o\ =\ \xi_0 + (\Theta(\beta_0)\eta\xi_0-\xi_0)P_{-3/2}
+ (\xi_0\eta\Theta(\beta_0)-\xi_0)P_{-1/2},
\end{equation}
where $P_{-3/2}$ ($P_{-1/2}$) is the projection operator onto the states with 
picture number $-3/2$ ($-1/2$).
We call the Hilbert space restricted by the constraint in Eq.~(\ref{open constraint}) 
the restricted Hilbert space and denote $\mathcal{H}_o^{res}$.
Note that $\mathcal{G}^o$ and $(\mathcal{G}^o)^{-1}$ satisfy
\begin{equation}
 \mathcal{G}^o(\mathcal{G}^o)^{-1}\mathcal{G}^0\ =\ \mathcal{G}^o,\qquad
(\mathcal{G}^o)^{-1}\mathcal{G}^o(\mathcal{G}^o)^{-1}\ =\ (\mathcal{G}^o)^{-1},\qquad
[Q,\mathcal{G}^o]\ =\ 0,
\label{G open}
\end{equation}
and thus $\mathcal{P}^o_{XY}$ is a projection operator that is compatible 
with the BRST cohomology: $Q\mathcal{P}^o_{XY}=\mathcal{P}^o_{XY}Q\mathcal{P}^o_{XY}$.
The open superstring field satisfying Eq.~(\ref{open constraint}) is expanded in the ghost zero-modes as
\begin{equation}
\Psi\ =\ (\phi_{NS} - c_0\psi_{NS}) + \left(\phi_R-\frac{1}{2}(\gamma_0+c_0G)\psi_R\right)\ 
\in \mathcal{H}_o^{res}.
\end{equation}
Natural symplectic form $\omega_s^o$ and $\Omega^o$ in $\mathcal{H}_o$ and $\mathcal{H}_o^{res}$,
respectively, are defined by using the BPZ inner product as
\begin{align}
 \omega_s^o(\Psi_1,\Psi_2)\ =&\ (-1)^{\textrm{deg}(\Psi_1)}\langle\Psi_1|\Psi_2\rangle,
\label{symp open s}\\
 \Omega^o(\Psi_1,\Psi_2)\ =&\ (-1)^{\textrm{deg}(\Psi_1)}\langle\Psi_1|(\mathcal{G}^o)^{-1}|\Psi_2\rangle,
\label{symp open res}
\end{align}
where deg$(\Psi)=1$ or $0$ if $\Psi$ is Grassmann even or odd, respectively. 
We also use a natural symplectic form $\omega_l^o$ in the large Hilbert space $\mathcal{H}_l^o$, 
which is similarly defined using the BPZ inner product in $\mathcal{H}_l^o$, 
and related to $\omega_s^o$ as $\omega_l^o(\xi_0\Psi_1,\Psi_2)=\omega_s^o(\Psi_1,\Psi_2)$
if $\Psi_1,\Psi_2\in\mathcal{H}_s^o$.

\subsection{Interaction with $A_\infty$-structure}

Open superstring interactions are described by the string products $M_n$ mapping $n$ open 
superstring fields to an open superstring field as
\begin{alignat}{3}
M_n\ :\qquad\qquad 
&\ (\mathcal{H}_o^{res})^{\otimes n} &\qquad &\ \longrightarrow&\qquad  &\ \mathcal{H}_o^{res},\qquad
(n\ge1),
\nonumber\\
&\ \ \ \  \text{\rotatebox[origin=c]{90}{$\in$}}&\qquad &\ &\qquad &\ \ \text{\rotatebox[origin=c]{90}{$\in$}}\\
\ \ \  \Psi_1&\otimes \ \cdots\otimes\Psi_n &\qquad &\ \longmapsto&\qquad  M_n(\Psi_1&,\ \cdots, \Psi_n).
\nonumber
\end{alignat}
We identify the one-string product as the open superstring BRST operator: $M_1=Q_o$. 
Note that the conditions
\begin{equation}
\mathcal{P}^o_{XY}M_n(\Psi_1,\cdots,\Psi_n)\ =\ M_n(\Psi_1,\cdots,\Psi_n)
\end{equation}
hold by definition. The multi-linear maps $M_n$ further satisfy the $A_\infty$ relations
\begin{equation}
 \sum_{m=0}^n\sum_{k=0}^{n-m}
(-1)^{\epsilon(1,k)}
M_{n-m+1}(\Psi_1,\cdot,\Psi_k,
M_{m+1}(\Psi_{k+1},\cdots,\Psi_{k+m+1}),\Psi_{k+m+2},\cdots,\Psi_n)\ =\ 0,
\label{A infinity}
\end{equation}
where $\epsilon(1,k)=\sum_{i=1}^k\textrm{deg}(\Psi_i)$, and cyclicity with respect to
the symplectic form $\Omega^o$,
\begin{equation}
 \Omega^o(\Psi_1,M_n(\Psi_2,\cdots,\Psi_{n+1}))\ =\ 
-(-1)^{\textrm{deg}(\Psi_1)}\Omega^o(M_n(\Psi_1,\cdots,\Psi_n),\Psi_{n+1}).
\label{cyclicity open}
\end{equation}
The linear maps satisfying (\ref{A infinity}) and (\ref{cyclicity open}) form
the cyclic $A_\infty$ algebra $(\mathcal{H}_o^{res},\Omega^o,\{M_m\})$.

Coalgebra representation allows us to describe these infinite number of relations 
of maps $M_n$ concisely \cite{Erler:2015uba}. The set of maps $\{M_n\}$ are represented by 
a degree-odd coderivation 
$\bd{M}=\sum_{n=1}^\infty \bd{M}_{n}$ 
acting on the tensor algebra
$\mathcal{TH}_o=\sum_{n=0}^\infty(\mathcal{H}_o^{res})^{\otimes n}$
as
\begin{equation}
\bd{M}\ =\ \sum_{n=1}^\infty\ \bd{M}_n\ 
=\ \sum_{n=1}^\infty\sum_{k,l=0}^\infty\left(\id^{\otimes k}\otimes M_n\otimes \id^{\otimes l}
\right)\pi^o_{k+n+l}, 
\end{equation}
where $\pi_m^o$ is the projection operator onto
$(\mathcal{H}_o^{res})^{\otimes m}\subset\mathcal{TH}_o$.
Then the $A_\infty$ relations in Eq.~(\ref{A infinity}) is concisely written as\footnote{
In this paper, $[\ ,\ ]$ denotes the graded commutator.}
\begin{equation}
 [\bd{M},\bd{M}]\ =\ 0.
\end{equation}
For open superstring field theory, the string interaction $M_n$ must be defined
for each combination of NS and R inputs
so that the picture number must be conserved\footnote{
Whether the output is NS or R string is determined by the space-time fermion number
conservation.}:
\begin{equation}
 \bd{M}_{n+1}\ =\ \sum_{p+r=n}\bd{M}_{p+r+1}^{(p)}|_{2r},
\end{equation}
where $p$ is the picture number that the the map itself has and
$2r$ is the Ramond number ($=$ number of Ramond inputs $-$ number of Ramond output).

The action with $A_\infty$ structure is given by
\begin{equation}
 I_o\ =\ \int_0^1dt\, \Omega^o\left(\Psi,\pi_1^o\bd{M}\left(\frac{1}{1-t\Psi}\right)\right),
\label{A action}
\end{equation}
where we introduce a real parameter $t\in[0,1]$ and the group-like element $\frac{1}{1-\Psi}$ defined by
\begin{equation}
 \frac{1}{1-\Psi}\ =\ \id_{\mathcal{TH}_o}+\sum_{n=1}^\infty\Psi^{\otimes n}.
\end{equation}
Here, $\id_{\mathcal{TH}_o}$ is the identity in $\mathcal{TH}_o$ satisfying 
$\id_{\mathcal{TH}_o}\otimes V = V = V\otimes\id_{\mathcal{TH}_o}$ for ${}^\forall V\in\mathcal{TH}_o$.
The arbitrary variation of $I_o$ is given by
\begin{equation}
 \delta I_o\ =\ \Omega^o\left(\delta\Psi,\pi_1^o\bd{M}\left(\frac{1}{1-\Psi}\right)\right),
\end{equation}
where we used the cyclicity in Eq.~(\ref{cyclicity open}).
We can show that the action in Eq.~(\ref{A action}) is invariant under the gauge transformation
\begin{equation}
 \delta_{\Lambda}\Psi\ =\ \pi_1^o\bd{M}\left(\frac{1}{1-\Psi}\otimes\Lambda\otimes\frac{1}{1-\Psi}\right),
\end{equation}
using the $A_\infty$ relation in Eq.~(\ref{A infinity}) and cyclicity in Eq.~(\ref{cyclicity open}):
\begin{align}
\delta_{\Lambda}I_o\ =&\ 
\Omega^o\left(\pi_1^o\bd{M}\left(\frac{1}{1-\Psi}\otimes\Lambda\otimes\frac{1}{1-\Psi}\right),
\pi_1^o\bd{M}\left(\frac{1}{1-\Psi}\right)\right)
\nonumber\\
=&\ \Omega^o\left(\Lambda, \pi_1^o\bd{M}\left(\frac{1}{1-\Psi}\otimes
\pi_1^o\bd{M}\left(\frac{1}{1-\Psi}\right)\otimes\frac{1}{1-\Psi}\right)\right)
\nonumber\\
=&\ \Omega^o\left(\Lambda,\pi_1^o\bd{MM}\left(\frac{1}{1-\Psi}\right)\right)\ =\ 0.
\end{align}

\subsection{Explicit construction of interactions}

The cyclic $A_\infty$ algebra $(\mathcal{H}_o^{res},\Omega^o,\bd{M})$ for open superstring field theory 
is constructed in two steps. First, we consider a cyclic $A_\infty$ algebra 
$(\mathcal{H}_l^o,\omega_l^o,\bd{Q}-\bd{\eta}+\bd{A})$.
A degree odd coderivation
\begin{equation}
\bd{A}=\sum_{p,r=0}^\infty\bd{A}_{p+r+1}^{(p)}|^{2r}\qquad
(\bd{A}_1^{(0)}\mid^0\equiv 0)
\label{A for A-inf}
\end{equation}
is defined respecting the cyclic Ramond number ($=$ number of Ramond inputs $+$ 
number of Ramond output) to make it easier to realize cyclicity\footnote{
Note that the cyclic Ramond number has the upper bound $p+2\ge r$. We consider
$\bd{A}_{p+r+1}^{(p)}\mid^{2r}\equiv 0 $ against the outside of this region.}.
This other $A_\infty$ algebra can be decomposed into two mutually commutative 
$A_\infty$ algebras $(\mathcal{H}_l,\bd{\mathcal{D}})$ and $(\mathcal{H}_l,\bd{\mathcal{C}})$ with
\begin{equation}
 \pi_1\bd{\mathcal{D}}\ =\ \pi_1\bd{Q} +\pi_1^0\bd{A},\qquad
 \pi_1\bd{\mathcal{C}}\ =\ \pi_1\bd{\eta} -\pi_1^1\bd{A}
\end{equation}
depending on the picture number deficit of the output.
The $A_\infty$ relation $[\bd{A},\bd{A}]=0$ can also be decomposed as
\begin{subequations} \label{A inf bare}
\begin{align}
 [\bd{Q},\bd{A}] + \frac{1}{2}[\bd{A},\bd{A}]^1\ =\ 0,\\
 [\bd{\eta},\bd{A}] - \frac{1}{2}[\bd{A},\bd{A}]^2\ =\ 0,
\end{align}
\end{subequations}
where the bracket with subscript $[\cdot,\cdot]^{1\,\textrm{or}\,2}$ is defined by 
projecting the intermediate state onto the NS or R state after taking the (graded) commutator.
The relation $[\cdot,\cdot]=[\cdot,\cdot]^1+[\cdot,\cdot]^2$ holds
since the intermediate state is either the NS state or R state. 
If such $A_\infty$ algebras are obtained, 
we can transform them by the cohomomorphism
\begin{equation}
 \hat{\bd{F}}^{-1}\ =\ \pi_1\id - \Xi^o\pi_1^1\bd{A}
\label{cohomo A}
\end{equation}
to the cyclic $A_\infty$ algebra of interest $(\mathcal{H}_o^{res},\Omega^o,\bd{M})$  
and a (trivial) $A_\infty$ algebra $(\mathcal{H}^o_l,\bd{\eta})$:
\begin{equation}
\pi_1\hat{\bd{F}}^{-1}\bd{\mathcal{D}}\hat{\bd{F}}\ =\ \pi_1\bd{Q} + \mathcal{G}^o\pi_1\bd{A}\hat{\bd{F}}\
\equiv\ \pi_1\bd{M},\qquad
\pi_1\hat{\bd{F}}^{-1}\bd{\mathcal{C}}\hat{\bd{F}}\ =\ \pi_1\bd{\eta}.
\end{equation}

We consider a generating function
\begin{equation}
 \bd{A}(s,t)\ =\ \sum_{p,m,r=0}^\infty s^mt^p\bd{A}_{m+p+r+1}^{(p)}|^{2r}\ \equiv\
\sum_{p=0}^\infty t^p \bd{A}^{(p)}(s)
\label{gen fun A}
\end{equation}
for constructing the $A_\infty$ algebra $(\mathcal{H}_l^o,\omega_l^o,\bd{Q}-\bd{\eta}+\bd{A})$
and extend the $A_\infty$ relations in Eqs.~(\ref{A inf bare}) to
\begin{subequations}\label{ext A infinity rel} 
\begin{align}
\bd{I}(s,t)\ \equiv\ [\bd{Q},\bd{A}(s,t)] + \frac{1}{2}[\bd{A}(s,t),\bd{A}(s,t)]_{\mathfrak{o}_1(s)}\ =&\ 0,\\
\bd{J}(s,t)\ \equiv\ [\bd{\eta},\bd{A}(s,t)] - \frac{1}{2}[\bd{A}(s,t),\bd{A}(s,t)]_{\mathfrak{o}_2(t)}\ =&\ 0,
\end{align}
\end{subequations}
by introducing parameters $s$ and $t$ counting the picture number deficit and the picture number,
respectively. 
Here, in Eqs.~(\ref{ext A infinity rel}), $\mathfrak{o}_1(s)\ =\ \pi^0+s\pi^1,\ \mathfrak{o}_2(t)\ =\ t\pi^1$ and
the bracket with subscript $[\cdot,\cdot ]_{\mathcal{O}}$ is another simple notation for $[\cdot,\cdot]^{1,2}$
and is defined by inserting the operator $\mathcal{O}$ into the intermediate state after taking (graded) commutation relation,
\begin{equation}
 \pi_1[\bd{D},\bd{D}']_{\mathcal{O}}\ =\ \sum_n\pi_1\Big( \bd{D}_n\big(\mathcal{O}\pi_1\bd{D}'\wedge\id_{n-1})
-(-1)^{DD'+\mathcal{O}(D+D')}\bd{D}'_n\big(\mathcal{O}\pi_1\bd{D}\wedge\id_{n-1}\big)\Big).
\end{equation}
At $(s,t)=(0,1)$, the generating function in Eq.~(\ref{gen fun A}) and the relations in Eqs.~(\ref{ext A infinity rel}) 
reduce to $\bd{A}(0,1)=\bd{A}$ and the $A_\infty$ relations in Eqs.~(\ref{A inf bare}), respectively.

Then, we can show that if $\bd{A}(s,t)$ satisfies the differential equations
\begin{subequations} \label{diff open}
 \begin{align}
 \partial_t\bd{A}(s,t)\ =&\ [\bd{Q},\bd{\mu}(s,t)]+[\bd{A}(s,t),\bd{\mu}(s,t)]_{\mathfrak{o}_1(s)}\\
\partial_s\bd{A}(s,t)\ =&\ [\bd{\eta},\bd{\mu}(s,t)] -[\bd{A}(s,t),\bd{\mu}(s,t)]_{\mathfrak{o}_2(t)},  
 \end{align}
\end{subequations}
with introducing the degree even coderivation
\begin{equation}
 \bd{\mu}(s,t)\ =\ \sum_{p,m,r=0}^\infty s^mt^p\bd{\mu}^{(p+1)}_{m+p+r+2}|^{2r}\ \equiv\
\sum_{p=0}^\infty t^p \bd{\mu}^{(p+1)}(s),
\end{equation}
the $t$ derivative of the left hand sides of the relations in Eqs.~(\ref{ext A infinity rel}) become
\begin{align}
 \partial_t\bd{I}(s,t)\ =&\ [\bd{I}(s,t),\bd{\mu}(s,t)]_{\mathfrak{o}_1(s)},\\
 \partial_t\bd{J}(s,t)\ =&\ [\bd{J}(s,t),\bd{\mu}(s,t)]_{\mathfrak{o}_1(s)}
-[\bd{I}(s,t),\bd{\mu}(s,t)]_{\mathfrak{o}_2(s)} - \partial_s\bd{I}(s,t).
\end{align}
Thus, if 
\begin{subequations} \label{A inf boson}
 \begin{align}
\bd{I}(s,0)\ =&\ [\bd{Q},\bd{A}(s,0)] + \frac{1}{2}[\bd{A}(s,0),\bd{A}(s,0)]_{\mathfrak{o}_1(s)}\ =\ 0,\\  
\bd{J}(s,0)\ =&\ [\bd{\eta},\bd{A}(s,0)]\ =\ 0,
 \end{align}
\end{subequations}
then $\bd{I}(s,t)=\bd{J}(s,t)=0$. 
However, the relations in Eqs.~(\ref{A inf boson}) are nothing less than those satisfied by 
the \textit{geometric} string products
constructed similarly to those for the bosonic $A_\infty$ algebra, $\bd{Q} + \bd{M}_B(s)$,
without any insertion:
\begin{equation}
\bd{M}_B(s)\ \equiv\ \sum_{m,r=0}^\infty s^m(\bd{M}_B)_{m+r+1}\mid^{2r}. 
\label{bosonic A}
\end{equation}
We can obtain the cyclic $A_\infty$ algebra, $(\mathcal{H}_l^o,\omega_l^o,\bd{Q}-\bd{\eta}+\bd{A})$
by recursively solving the differential equations in Eqs.~(\ref{diff open}), or equivalently, 
\begin{subequations} 
 \begin{align}
  (p+1)\bd{A}^{(p+1)}(s)\ =&\ [\bd{Q},\bd{\mu}^{(p+1)}(s)]
+ \sum_{q=0}^p [\bd{A}^{(p-q)}(s),\bd{\mu}^{(q+1)}(s)]_{\mathfrak{o}_1(s)}\,,
\label{rec1}\\
[\bd{\eta},\bd{\mu}^{(p+1)}(s)] =&\ 
\partial_s\bd{A}^{(p)}(s) + \sum_{q=0}^{p-1}[\bd{A}^{(p-q-1)}(s),\bd{\mu}^{(q+1)}(s)]^2\,.
\label{rec2}
 \end{align}
\end{subequations}
The second equation at $p=0$ can be solved for $\bd{\mu}^{(1)}(s)$ as
\begin{equation}
 \bd{\mu}^{(1)}(s)\ =\ \xi_0^o\circ\partial_s\bd{M}_B(s)
\label{mu1}
\end{equation}
under the initial condition $\bd{A}(s,0)=\bd{A}^{(0)}(s)=\bd{M_B}(s)$\,, where
$\xi_0\circ$ is the operation defined on general coderivation $\bd{A}=\sum_{n=0}^\infty\bd{A}_{n+2}$ by
\begin{subequations} 
\begin{align}
 \xi_0^o\circ \bd{A}\ =&\ 
\sum_{n,k,l=0}^\infty\left(\id^{\otimes k}\otimes
\left(\xi_0^o\circ A_{n+2}\right)\otimes\id^{\otimes l}
\right)\pi^o_{k+l+n+2}\,,\\
\xi_0^o\circ A_{n+2}\ =&\
\frac{1}{n+3}\left(\xi_0^o A_{n+2} - (-1)^{\textrm{deg(A)}} \sum_{m=0}^{n+1}
A_{n+2}\left(\id^{\otimes (n-m+1)}\otimes\xi_0^o\otimes\id^{\otimes m}\right)\right)\,.
\end{align}
\end{subequations}
Substituting (\ref{mu1}) into (\ref{rec1}) at $p=0$, we obtain $\bd{A}^{(1)}(s)$\,.
Repeating the procedure, we can recursively obtain $\bd{\mu}^{(p+1)}(s)$ and $\bd{A}^{(n+1)}(s)$
from Eqs.~(\ref{rec2}) and $(\ref{rec1})$, respectively.\footnote{
Note that the part of $\bd{A}^{(p+1)}(s)$ with a fixed number of inputs contains
only a finite number of terms.
We can calculate any $\bd{A}_{p+r+1}^{(p)}\mid^{2r}$ with a finite procedure.} 
Then finally, the cohomomorphism in Eq.~(\ref{cohomo A}) 
gives the cyclic $A_\infty$ algebra $(\mathcal{H}^{res}_o,\Omega^o,\bd{M})$.

\sectiono{Closed superstring field theory with $L_\infty$-structure}\label{B}

Similarly to the open superstring field theory, closed (type II) superstring field theory 
is constructed based on the $L_\infty$ algebra structure. We next summarize it 
in this section.

\subsection{Closed superstring field}

The first-quantized Hilbert space, $\mathcal{H}_c$, of type II (closed)
superstring is composed of four sectors: 
$ \mathcal{H}_c = \mathcal{H}_{\NSNS} + \mathcal{H}_{\RNS} + \mathcal{H}_{\NSR} + \mathcal{H}_{\RR}$.
Correspondingly, the type II superstring field $\Phi$ has four components,
$\Phi = \Phi_{\NSNS} + \Phi_{\RNS} + \Phi_{\NSR} + \Phi_{\RR}$,
all of which are Grassmann even and have ghost number $2$.
The components $\Phi_{\NSNS}$ and $\Phi_{\RR}$ have picture numbers $(-1,-1)$ and $(-1/2,-1/2)$, 
respectively and represent space-time bosons. The components $\Phi_{\RNS}$ and $\Phi_{\NSR}$ 
have picture numbers $(-1/2,-1)$ and $(-1,-1/2)$ and represent space-time fermions. 
We impose it closed string constraints
\begin{equation}
 b_0^-\Phi\ =\ L_0^-\Phi\ =\ 0,
\end{equation}
and also an extra constraint
\begin{equation}
\mathcal{P}_{XY}^c\Phi\ =\ \Phi,\qquad \mathcal{P}_{XY}^c\ =\ \mathcal{G}^c(\mathcal{G}^c)^{-1},
\label{closed constraint}
\end{equation}
with
\begin{align}
 \mathcal{G}^c\ =&\ 
\pi^{(0,0)} + X^c \pi^{(1,0)} + \bar{X}^c \pi^{(0,1)} + X^c\bar{X}^c\pi^{(1,1)},\\
 (\mathcal{G}^c)^{-1}\ =&\ 
\pi^{(0,0)} + Y^c \pi^{(1,0)} + \bar{Y}^c \pi^{(0,1)} + Y^c\bar{Y}^c\pi^{(1,1)},
\end{align} 
where $\pi^{(0,0)}$, $\pi^{(1,0)}$, $\pi^{(0,1)}$, and $\pi^{(1,1)}$, are the projection
operators onto the NS-NS, R-NS, NS-R, and R-R components respectively:
 $\pi^{(0,0)}\Phi=\Phi_{\NSNS}$, $\pi^{(1,0)}\Phi=\Phi_{\RNS}$, 
$\pi^{(0,1)}\Phi=\Phi_{\NSR}$, and $\pi^{(1,1)}\Phi=\Phi_{\RR}$.
The PCO $X^c$ $(\bar{X}^c)$ and and its inverse $Y^c$ $(\bar{Y}^c)$ are defined by
\begin{align}
 X^c\ =&\ -\delta(\beta_0)G + \frac{1}{2}(\gamma_0\delta(\beta_0)+\delta(\beta_0)\gamma_0)b^+_0,\qquad 
 Y^c\ =\ - 2\frac{G}{L^+_0}\delta(\gamma_0),\\
 \bar{X}^c\ =&\ -\delta(\bar{\beta}_0)\bar{G} + \frac{1}{2}
(\bar{\gamma}_0\delta(\bar{\beta}_0)+\delta(\bar{\beta}_0)\bar{\gamma}_0)b^+_0,\qquad 
 \bar{Y}^c\ =\ - 2\frac{\bar{G}}{L^+_0}\delta(\bar{\gamma}_0).
\end{align}
The PCOs $X^c$ and $\bar{X}^c$ are BRST exact in the large Hilbert space:
\begin{alignat}{2}
 X^c\ =&\ [Q,\Xi^c],\qquad& 
\Xi^c\ =&\ \xi_0 + (\Theta(\beta_0)\eta\xi_0-\xi_0)P_{-3/2}+(\xi_0\eta\Theta(\beta_0)-\xi_0)P_{-1/2},\\
\bar{X}^c\ =&\ [Q,\bar{\Xi}^c],\qquad&
\bar{\Xi}^c\ =&\ \bar{\xi}_0 + (\Theta(\bar{\beta}_0)\bar{\eta}\bar{\xi}_0-\bar{\xi}_0)\bar{P}_{-3/2}
+(\bar{\xi}_0\bar{\eta}\Theta(\bar{\beta}_0)-\bar{\xi}_0)\bar{P}_{-1/2},
\end{alignat}
where $P_{-3/2}$ and $P_{-1/2}$ ($\bar{P}_{-3/2}$ and $\bar{P}_{-1/2}$)
are the projectors onto the states with the left-moving (right-moving) 
picture numbers $-3/2$ and $-1/2$, respectively. 
We denote the restricted Hilbert space of type II superstring as $\mathcal{H}_c^{res}$.
Similarly to the relations in Eq.~(\ref{G open}) for the open superstring, 
$\mathcal{G}^c$ and $(\mathcal{G}^c)^{-1}$ satisfy the relations
\begin{align}
\mathcal{G}^c (\mathcal{G}^c)^{-1} \mathcal{G}^c \ =\ \mathcal{G}^c,\qquad 
(\mathcal{G}^c)^{-1}\mathcal{G}^c (\mathcal{G}^c)^{-1}\ =\ (\mathcal{G}^c)^{-1},\qquad
[Q,\mathcal{G}^c]\ =\ 0,
\end{align}
and thus, $\mathcal{G}^c(\mathcal{G}^c)^{-1}$ is a projector that is compatible with 
the BRST cohomology: $Q\mathcal{P}_{XY}^c=\mathcal{P}_{XY}^cQ\mathcal{P}_{XY}^c$.
The type II superstring field satisfying the constraint in Eq.~(\ref{closed constraint})
is expanded in the ghost zero-modes as
\begin{align}
 \Phi\ =&\ (\phi_{\NSNS}-c_0^+\psi_{\NSNS}) 
+ \left(\phi_{\RR}-\frac{1}{2}(\gamma_0\bar{G}-\bar{\gamma}_0G
+2c_0^+G\bar{G})\psi_{\RR}\right) 
\nonumber\\
&\
+ \left(\phi_{\RNS}-\frac{1}{2}(\gamma_0+2c_0^+G)\psi_{\RNS}\right)
+ \left(\phi_{\NSR}-\frac{1}{2}(\bar{\gamma}_0+2c_0^+\bar{G})\psi_{\NSR}\right)
\in\mathcal{H}_c^{res}.
\end{align}
Natural symplectic forms $\omega_s^c$ and $\Omega^c$ in $\mathcal{H}_c$ and $\mathcal{H}_c^{res}$,
respectively, are defined by using the BPZ inner product as
\begin{align}
 \omega_s^c(\Phi_1,\Phi_2)\ =&\ (-1)^{\Phi_1}\langle\Phi_1|c_0^-|\Phi_2\rangle,\\
 \Omega^c(\Phi_1,\Phi_2)\ =&\ (-1)^{\Phi_1}\langle\Phi_1|c_0^-(\mathcal{G}^c)^{-1}|\Phi_2\rangle.
\end{align}
Natural symplectic form $\omega_l^c$ in the large Hilbert space $\mathcal{H}_l^c$ is similarly
defined by using the BPZ inner product in $\mathcal{H}_l^c$, and related to $\omega_s^c$ as
$\omega_l^c(\xi_0\bar{\xi}_0\Phi_1,\Phi_2)=\omega_s^c(\Phi_1,\Phi_2)$ if $\Phi_1,\ 
\Phi_2\in\mathcal{H}_s^c$.

\subsection{Interaction with $L_\infty$-structure}

Type II superstring interactions are descried by the string products $L_n$ that map
$n$ closed superstring fields to a closed superstring field as
\begin{alignat}{3}\label{L maps}
 L_n\ :\qquad\qquad 
&\ (\mathcal{H}_c^{res})^{\wedge n} &\qquad &\ \longrightarrow&\qquad  &\ \mathcal{H}_c^{res},\qquad
(n\ge1),
\nonumber\\
&\ \ \ \  \text{\rotatebox[origin=c]{90}{$\in$}}&\qquad &\ &\qquad &\ \ \text{\rotatebox[origin=c]{90}{$\in$}}\\
\ \ \  \Phi_1&\wedge \ \cdots\wedge\Phi_n &\qquad &\ \longmapsto&\qquad  L_n(\Phi_1& ,\ \cdots, \Phi_n),
\nonumber
\end{alignat}
where $\Phi_1\wedge\cdots\wedge\Phi_n$ is the symmetrized tensor product defined by
\begin{equation}
 \Phi_1\wedge\cdots\wedge\Phi_n\ =\ \sum_{\sigma}\Phi_{\sigma(1)}\otimes\cdots\otimes\Phi_{\sigma(n)},
\end{equation}
We identify the one-string product as the closed superstring BRST operator: $L_1=Q_c$.
By definition, these products must satisfy
\begin{align}
&\ b_0^-L_n(\Phi_1,\cdots,\Phi_n)\ =\  L_0^-L_n(\Phi_1,\cdots,\Phi_n)\ =\ 0,\\
&\ \mathcal{P}_{XY}^cL_n(\Phi_1,\cdots,\Phi_n)\ =\ L_n(\Phi_1,\cdots,\Phi_n).
\end{align}
We further impose the $L_\infty$ relations
\begin{equation}
\sum_{\sigma} \sum_{m=1}^n
(-1)^{\epsilon(\sigma)}
\frac{1}{m!(n-m)!}
L_{n-m+1}(L_m(\Phi_{\sigma(1)},\cdots,\Phi_{\sigma(m)}),\Phi_{\sigma(m+1)},\cdots,\Phi_{\sigma(n)})\ =\ 0,
\label{L infinity}
\end{equation}
and cyclicity 
\begin{equation}
 \Omega^c(\Phi_1,L_n(\Phi_2,\cdots,\Phi_{n+1}))\ =\ 
-(-1)^{|\Phi_1|}\Omega^c(L_n(\Phi_1,\cdots,\Phi_n),\Phi_{n+1}).
\label{cyclicity closed}
\end{equation}
The linear maps satisfying Eqs.~(\ref{L infinity}) and (\ref{cyclicity closed}) form
the cyclic $L_\infty$ algebra $(\mathcal{H}_c^{res},\Omega^c,\{L_m\})$.

The linear maps in Eq.~(\ref{L maps}) are also represented by a degree-odd coderivation
$\bd{L}=\sum_{n=1}^\infty\bd{L}_n$ acting on the symmetrized tensor algebra
$\mathcal{SH}_c=\sum_{n=0}^\infty(\mathcal{H}_c^{res})^{\wedge n}$ as
\begin{equation}
 \bd{L}\ =\ \sum_{n=1}^\infty\ \bd{L}_n\
=\ \sum_{n=1}^\infty\sum_{m=0}^\infty 
(L_n\wedge\id_m)\pi_{n+m}^c.
\end{equation}
with 
$
\id_m\ =\ \frac{1}{m!}\, \id^{\wedge m}
=\ \id^{\otimes m}, 
$
where $\pi_m^c$ is the projection operator onto
$(\mathcal{H}_c^{res})^{\wedge m}\subset\mathcal{SH}_c$.

Then, the $L_\infty$ relations in Eq.~(\ref{L infinity}) is written as
\begin{equation}
 [\bd{L},\bd{L}]\ =\ 0.
\end{equation}

The string interaction $L_n$ 
of the type II superstring field theory must be defined
for each combination of NS-NS, R-NS, NS-R and R-R inputs so that the picture numbers
of left- and right-moving sectors must be conserved separately:
\begin{equation}
 \bd{L}_{n+1}\ =\ \sum_{p+r=n}\sum_{\bar{p}+\bar{m}=n}
\bd{L}^{(p,\bar{p})}_{p+r+1,\bar{p}+\bar{m}+1}\mid_{(2r,2\bar{r})},
\end{equation}
where we used the diagonal matrix representation $\bd{L}_{n,m}=\delta_{n,m}\bd{L}_n$.
The superscript $p$ ($\bar{p}$) is the left-moving (right-moving) 
picture number that the map itself has, 
and the subscript $2r$ ($2\bar{r}$) is the left-moving (right-moving) Ramond number.

Introducing a real parameter $t\in[0,1]$, 
the action with $L_\infty$ structure is given by
\begin{equation}
 I_c\ =\ \int^1_0dt\,\Omega^c(\Phi,\pi_1^c\bd{L}(e^{\wedge t\Phi})),
\label{L action}
\end{equation}
with the group-like element 
\begin{equation}
 e^{\wedge\Phi}\ =\ \id_{\mathcal{SH}_c} + \sum_{n=1}^\infty\frac{1}{n!}\Phi^{\wedge n},
\end{equation}
where $\id_{\mathcal{SH}_c}$ is the identity in $\mathcal{SH}_c$ that satisfies
$\id_{\mathcal{SH}_c}\wedge V=V$ for ${}^{\forall}V\in\mathcal{SH}_c$. 
The arbitrary variation of $I_c$ is given by
\begin{equation}
 \delta I_c\ =\ \Omega^c(\delta\Phi,\pi_1\bd{L}(e^{\wedge\Phi})).
\end{equation}
We can show the action in Eq.~(\ref{L action}) is invariant under the gauge transformation
\begin{equation}
 \delta_{\Lambda}\Phi\ =\ \pi_1^c\bd{L}(e^{\wedge\Phi}\wedge\Lambda),
\end{equation}
using the $L_\infty$ relation in Eq.~(\ref{L infinity}) and cyclicity in Eq.~(\ref{cyclicity closed}):
\begin{align}
 \delta_{\Lambda} I_c\ =&\ \Omega^c\left(\pi_1^c\bd{L}(e^{\wedge\Phi}\wedge\Lambda),
\pi_1^c\bd{L}(e^{\wedge\Phi})\right)
\nonumber\\
=&\ \Omega^c\left(\Lambda,\pi_1^c\bd{L}\left(e^{\wedge\Phi}\wedge\pi_1^c\bd{L}(e^{\wedge\Phi})\right)\right)
\nonumber\\
=&\ \Omega^c\left(\Lambda,\pi_1^c\bd{L}^2(e^{\wedge\Phi})\right)\ =\ 0.
\end{align}

\subsection{Explicit construction of interactions}

The cyclic $L_\infty$ algebra $(\mathcal{H}_c^{res},\Omega^c,\bd{L})$ is constructed
in two steps. We consider first an $L_\infty$ algebra 
$(\mathcal{H}_l^c,\bd{\mathcal{O}})$ with
\begin{equation}
 \pi_1\bd{\mathcal{O}}\ =\ \pi_1(\bd{Q} - \bd{\eta} - \bar{\bd{\eta}}+\bd{B})
-\left(1-\frac{1}{2}(X+\bar{X})\right)\pi_1^{(1,1)}\bd{B}.
\end{equation} 
introducing the degree odd coderivation 
\begin{equation}
\bd{B}=\sum_{p,r=0}^\infty\sum_{\bar{p},\bar{r}=0}^\infty
\bd{B}_{p+r+1,\bar{p}+\bar{r}+1}^{(p,\bar{p})}\mid^{(2r,2\bar{r})}. 
\label{B for L-inf}
\end{equation}
This $L_\infty$ algebra is equivalent to three mutually commutative $L_\infty$
algebras $(\mathcal{H}_l,\bd{\mathcal{D}})$, 
$(\mathcal{H}_l,\bd{\mathcal{C}})$, and $(\mathcal{H}_l,\bar{\bd{\mathcal{C}}})$
with
\begin{align}
 \pi_1\bd{\mathcal{D}}\ =&\ \pi_1\bd{Q}+\pi_1^{(0,0)}\bd{B},\\
 \pi_1\bd{\mathcal{C}}\ =&\ \pi_1\bd{\eta}-\left(\pi_1^{(1,0)}
+\frac{1}{2}\bar{X}\pi_1^{(1,1)}\right)\bd{B},\qquad
 \pi_1\bar{\bd{\mathcal{C}}}\ =\ \pi_1\bar{\bd{\eta}}-\left(\pi_1^{(0,1)}
+\frac{1}{2}X\pi_1^{(1,1)}\right)\bd{B}.
\end{align}
decomposed according to the picture number deficit.
Then, the $L_\infty$ relations are written as
\begin{subequations} \label{L inf bare}
 \begin{align}
 [\bd{Q},\bd{B}] + \frac{1}{2}[\bd{B},\bd{B}]^{11}\ =&\ 0,\\
 [\bd{\eta},\bd{B}] -\frac{1}{2}[\bd{B},\bd{B}]^{21} -\frac{1}{4}[\bd{B},\bd{B}]_{\bar{X}}^{22}\ =&\ 0,\\ 
 [\bar{\bd{\eta}},\bd{B}] -\frac{1}{2}[\bd{B},\bd{B}]^{12} -\frac{1}{4}[\bd{B},\bd{B}]_{X}^{22}\ =&\ 0.
 \end{align}
\end{subequations}
Here, the bracket $[\cdot,\cdot]^{11,21,12,22}$ is defined by projecting the intermediate
state of the (graded) commutator to the NS-NS, R-NS, NS-R, or R-R state. We also define the bracket
$[\cdot,\cdot]^{22}_{X\,\textrm{or}\,\bar{X}}$ by further inserting $X$ or $\bar{X}$ at
the intermediate R-R state.
If such $L_\infty$ algebras are found, we transform them by 
cohomomorphism
\begin{equation}
 \pi_1\hat{\bd{F}}^{-1}\ =\ \pi_1\id 
-\left(\Xi\pi_1^{(1,0)}+\bar{\Xi}\pi^{(0,1)}_1
+\frac{1}{2}(\Xi\bar{X}+X\bar{\Xi})\pi_1^{(1,1)}\right)\bd{B}
\label{cohomo L}
\end{equation}
to the cyclic $L_\infty$ algebra $(\mathcal{H}_c^{res},\Omega^c,\bd{L})$
and two (trivial) $L_\infty$ algebras $(\mathcal{H}_l^c,\bd{\eta})$
and $(\mathcal{H}_l^c,\bar{\bd{\eta}})$ as
\begin{align}
\pi_1\hat{\bd{F}}^{-1}\bd{\mathcal{D}}\hat{\bd{F}}\ =&\
\pi_1\bd{Q}+\mathcal{G}^c\pi_1\bd{B}\hat{\bd{F}}\ \equiv \pi_1\bd{L},
\label{final L}\\ 
\pi_1\hat{\bd{F}}^{-1}\bd{\mathcal{C}}\hat{\bd{F}}\ =&\ \pi_1\bd{\eta},\qquad
\pi_1\hat{\bd{F}}^{-1}\bar{\bd{\mathcal{C}}}\hat{\bd{F}}\ =\ \pi_1\bar{\bd{\eta}}.
\end{align}
Note that the $L_\infty$ algebra $(\mathcal{H}_l^c,\bd{\mathcal{O}})$ 
is not cyclic with respect to $\omega_l^c$
unlike the open superstring case. However, we can show,
in a similar way given in the Appendix C of Ref.~\cite{Kunitomo:2019glq},
that the $\bd{L}$ in Eq.~(\ref{final L}) is cyclic with respect to $\Omega^c$
if $\bd{B}$ is cyclic with respect to $\omega_l^c$.

In the next step, we consider a generating function
\begin{equation}
 \bd{B}\sst\ =\ \sum_{m,p,r=0}^\infty\sum_{\bar{m},\bar{p},\bar{r}=0}^\infty
s^m\bar{s}^{\bar{m}}t^{p+\bar{p}}\bd{B}^{(p,\bar{p})}_{m+p+r+1,\bar{m}+\bar{p}+\bar{r}+1}\mid^{(2r,2\bar{r})}
\label{gen fun L}
\end{equation}
and extend the $L_\infty$ relations in Eq.~(\ref{L inf bare})  to
\begin{subequations} \label{ext L infinity rel} 
 \begin{align}
\bd{I}\sst\ \equiv&\ [\bd{Q},\bd{B}\sst] + \frac{1}{2}[\bd{B}\sst,\bd{B}\sst]_{\mathfrak{c}_1\sst}\ =\ 0,\\
\bd{J}\sst\ \equiv&\ [\bd{\eta},\bd{B}\sst] -\frac{1}{2}[\bd{B}\sst,\bd{B}\sst]_{\mathfrak{c}_2(t)}\ =\ 0,\\
\bar{\bd{J}}\sst\ \equiv&\ [\bar{\bd{\eta}},\bd{B}\sst] -\frac{1}{2}[\bd{B}\sst,\bd{B}\sst]_{\bar{\mathfrak{c}}_2(t)}\ =\ 0,
\end{align}
\end{subequations}
for constructing the $L_\infty$ algebra $(\mathcal{H}_l^c,\bd{\mathcal{O}})$.
The parameters $s$, $\bar{s}$, and $t$ counting the left-moving picture number deficit,
right-moving picture number deficit, and the total picture number, respectively.
The bracket with subscript is defined by inserting
\begin{align}
 \mathfrak{c}_1\sst\ =&\ \pi^{(0,0)} 
+s\pi^{(1,0)}+\bar{s}\pi^{(0,1)}
+\left(s\bar{s}+t(s\bar{X}+\bar{s}X)\right)\pi^{(1,1)},\\
\mathfrak{c}_2(t)\ =&\ t\pi^{(1,0)}+\frac{t^2}{2}\bar{X}\pi^{(1,1)},\qquad
\bar{\mathfrak{c}}_2(t)\ =\ t\pi^{(0,1)}+\frac{t^2}{2}X\pi^{(1,1)},
\end{align}
at the intermediate state. 
At $(s,\bar{s},t)=(0,0,1)$, the generating function in Eq.~(\ref{gen fun L}) and the relations 
in Eq.~(\ref{ext L infinity rel}) 
reduce to $\bd{B}(0,0,1)=\bd{B}$ and the $L_\infty$ relations in Eqs.~(\ref{L inf bare}), respectively.

We can show that if $\bd{B}\sst$ satisfies
the differential equations
\begin{subequations}\label{diff closed}
 \begin{align}
 \partial_t\bd{B}\sst\ =&\ [\bd{Q},(\bd{\lambda}+\bar{\bd{\lambda}})\sst]
\nonumber\\
&\
+ [\bd{B}\sst,(\bd{\lambda}+\bar{\bd{\lambda}})\sst]_{\mathfrak{c}_1\sst}
+\frac{1}{2}[\bd{B}\sst,\bd{B}\sst]_{\mathfrak{d}(s,\bar{s})}\,
\label{diff closed t}\\
\partial_s\bd{B}\sst\ =&\ [\bd{\eta},\bd{\lambda}\sst]
- [\bd{B}\sst,(\bd{\lambda}+\bar{\bd{\lambda}})\sst]_{\mathfrak{c}_2(t)},\\
\partial_{\bar{s}}\bd{B}\sst\ =&\ [\bar{\bd{\eta}},\bar{\bd{\lambda}}\sst]
- [\bd{B}\sst,(\bd{\lambda}+\bar{\bd{\lambda}})\sst]_{\bar{\mathfrak{c}}_2(t)},
\end{align}
\end{subequations}
and $[\bar{\bd{\eta}},\bd{\lambda}\ssb]=[\bd{\eta},\bar{\bd{\lambda}}\ssb]=0$ 
with $\mathfrak{d}(s,\bar{s})=(s\bar{\Xi}+\bar{s}\Xi)\pi^{(1,1)}$ and
the degree even coderivations
\begin{align}
 \bd{\lambda}\sst\ =&\ 
\sum_{m,p,r=0}^\infty\sum_{\bar{m},\bar{p},\bar{r}=0}^\infty s^m
\bar{s}^{\bar{m}}t^{p+\bar{p}}
\bd{\lambda}^{(p+1,\bar{p})}_{m+p+r+2,\bar{m}+\bar{p}+\bar{r}+1}\mid^{(2r,2\bar{r})},\\
 \bar{\bd{\lambda}}\sst\ =&\ \sum_{m,p,r=0}^\infty\sum_{\bar{m},\bar{p},\bar{r}=0}^\infty
s^m\bar{s}^{\bar{m}}t^{p+\bar{p}}
\bar{\bd{\lambda}}^{(p,\bar{p}+1)}_{m+p+r+1,\bar{m}+\bar{p}+\bar{r}+2}\mid^{(2r,2\bar{r})},
\end{align}
the $t$ derivative of the left hand sides of the relations in Eqs.~(\ref{ext L infinity rel}) become
\begin{subequations} 
 \begin{align}
 \partial_t\bd{I}\sst\ =&\ [\bd{I}\sst,(\bd{\lambda}+\bar{\bd{\lambda}})\sst]_{\mathfrak{c}_1\sst}
 + [\bd{I}\sst,\bd{B}\sst]_{\mathfrak{d}(s,\bar{s})},\\
\partial_t\bd{J}\sst\ =&\ [\bd{J}\sst,(\bd{\lambda}+\bar{\bd{\lambda}})\sst]_{\mathfrak{c_1}\sst} 
+ [\bd{J}\sst,\bd{B}\sst]_{\mathfrak{d}(s,\bar{s})} 
\nonumber\\
&\
-\partial_s\bd{I}\sst 
- [\bd{I}\sst,(\bd{\lambda}+\bar{\bd{\lambda}})\sst]_{\mathfrak{c}_2(t)},\\  
\partial_t\bar{\bd{J}}\sst\ =&\ [\bar{\bd{J}}\sst,(\bd{\lambda}+\bar{\bd{\lambda}})\sst]_{\mathfrak{c_1}\sst} 
+ [\bar{\bd{J}}\sst,\bd{B}\sst]_{\mathfrak{d}(s,\bar{s})} 
\nonumber\\
&\
-\partial_{\bar{s}}\bd{I}\sst - [\bd{I}\sst,(\bd{\lambda}+\bar{\bd{\lambda}})\sst]_{\bar{\mathfrak{c}}_2(t)}.
 \end{align}
\end{subequations}
They imply if
\begin{subequations} \label{L inf boson}
\begin{align}
 \bd{I}(s,\bar{s},0)\ =&\ [\bd{Q},\bd{B}(s,\bar{s},0)] 
+ \frac{1}{2}[\bd{B}(s,\bar{s},0),\bd{B}(s,\bar{s},0)]_{\mathfrak{c}_1(s,\bar{s},0)}\ =\ 0,\\
\bd{J}(s,\bar{s},0)\ =&\ [\bd{\eta},\bd{B}(s,\bar{s},0)]\ =\ 0,\\
\bar{\bd{J}}(s,\bar{s},0)\ =&\ [\bar{\bd{\eta}},\bd{B}(s,\bar{s},0)]\ =\ 0,
\end{align}
\end{subequations}
then $\bd{I}\sst=\bd{J}\sst=\bar{\bd{J}}\sst=0$.
The relations in Eqs.~(\ref{L inf boson}) are those satisfied by the geometric string product
without any insertion, which can be constructed similarly to that for the bosonic $L_\infty$ algebra, 
$\bd{Q}+\bd{L}_B(s,\bar{s})$:
\begin{equation}
\bd{L}_B(s,\bar{s})\ \equiv\
\sum_{m,r=0}^\infty\sum_{\bar{m},\bar{r}=0}^\infty
s^m\bar{s}^{\bar{m}}(\bd{L}_B)_{m+r+1,\bar{m}+\bar{r}+1}\mid^{(2r,2\bar{r})}.
\label{bosonic L}
\end{equation}
Similar to the open-superstring case in the previous section,
we can obtain the $L_\infty$ algebra $(\mathcal{H}_l^c,\bd{\mathcal{O}})$ by solving
the differential equations in Eqs.~(\ref{diff closed}) with the initial condition
\begin{equation}
 \bd{B}(s,\bar{s},0)=\bd{L}_B(s,\bar{s})\,.
\label{closed ini}
\end{equation}
The concrete procedures are slightly complicated, so Appendix~\ref{App A0} 
shows the lower-order results.
The cohomomorphism in Eq.~(\ref{cohomo L}) gives the cyclic $L_\infty$ algebra $(\mathcal{H}_c^{res},\Omega^c,\bd{L})$
if we choose the solution $\bd{B}$ to be cyclic with respect to $\omega_l^c$.

\sectiono{Open-closed superstring field theory with OCHA-structure}\label{C}

Now, we are ready to discuss OCHA, the main subject of this paper. 
In this section, we first see what OCHA is and how it is realized in the superstring 
field theory and then give a prescription to construct them explicitly.

\subsection{Interaction with OCHA structure}

We define classical interactions among the open and closed (type II) superstrings mixed
system by the vertices described by the following two kinds of surfaces:
\begin{itemize}
 \item A sphere with $n\, (\ge3)$ closed-superstring punctures;
 \item A disk with $n\,(\ge0)$ closed-superstring punctures on the bulk 
and $l+1\,(\ge1)$ open-superstring punctures on the boundary with $n+l\ge1$.
\end{itemize}

We can identify the former as the linear maps $\{L_n\}$ given in the previous section, 
which form the cyclic $L_\infty$ algebra $(\mathcal{H}_c^{res},\Omega^c,\{L_n\})$.
The latter includes both the open-superstring interactions ($n=0$) 
and interactions between open and closed superstrings ($n>0$) and is described by 
the string products $N_{n,l}$ that maps
$n$ closed-superstring fields and $l$ open-superstring fields to an open-superstring field:
\begin{alignat}{3}\label{linear map n}
 N_{n,l}\ :\qquad
&\ (\mathcal{H}_c^{res})^{\wedge n}\otimes(\mathcal{H}_o^{res})^{\oplus l} 
    &\ &\longrightarrow& &\ \mathcal{H}_o^{res},\quad
(n,l\ge0,\,n+l>0),
\nonumber\\
&\ \ \ \ \ \ \ \ \ \ \ \ \ 
\text{\rotatebox[origin=c]{90}{$\in$}} &\ & &\ &\ \text{\rotatebox[origin=c]{90}{$\in$}}\\
\ \ \  (\Phi_1&\wedge \ \cdots\wedge\Phi_n)\otimes(\Psi_1\otimes\cdots\otimes\Psi_l)
  &\ &\longmapsto &\  N_{n,l}(\Phi_1&,\ \cdots, \Phi_n ; \Psi_1,\cdots,\Psi_l),
\nonumber
\end{alignat}
with the identification $N_{0,l}=M_l$.
By definition, the condition
\begin{equation}
\mathcal{P}_{XY}^oN_{n,l}(\Phi_1,\cdots,\Phi_n;\Psi_1,\cdots,\Psi_l)\ =\ 
N_{n,l}(\Phi_1,\cdots,\Phi_n;\Psi_1,\cdots,\Psi_l)
\end{equation}
holds. The linear maps $\{L_n,N_{n,l}\}$ satisfying the OCHA relation
\begin{align}\label{OCHA rel}
  0\ =&\ \sum_{\sigma}\sum_{m=1}^n(-1)^{\epsilon(\sigma)}\frac{1}{m!(n-m)!}N_{n-m+1,l}(L_{m}(\Phi_{\sigma(1)},\cdots,
\Phi_{\sigma(m)}),\Phi_{\sigma(m+1)},\cdots,\Phi_{\sigma(n)};\Psi_1,\cdots,\Psi_l)
\nonumber
\\
&\
+ \sum_{\sigma}\sum_{m=0}^n\sum_{j=0}^{l}\sum_{i=0}^{l-j}(-1)^{\mu_{m,i}(\sigma)}
\frac{1}{m!(n-m)!}
N_{m,l-j+1}(\Phi_{\sigma(1)},\cdots,\Phi_{\sigma(m)};
\nonumber\\
&\hspace{20mm}
\Psi_1,\cdots,\Psi_i,
N_{n-m,j}(\Phi_{\sigma(m+1)},\cdots,\Phi_{\sigma(n)};\Psi_{i+1},\cdots,\Psi_{i+j}),\Psi_{i+j+1},\cdots,\Psi_l),
\end{align}
and the cyclicity condition
\begin{align}
&\ \Omega^o(\Psi_1,N_{n,l}(\Phi_1,\cdots,\Phi_n;\Psi_2,\cdots,\Psi_{l+1}))
\nonumber\\ 
&\hspace{2cm}
=\ 
-(-1)^{\textrm{deg}(\Psi_1)(|\Phi_1|+\cdots+|\Phi_l|+1)}
 \Omega^o(N_{n,l}(\Phi_1,\cdots,\Phi_n;\Psi_1,\cdots,\Psi_l),\Psi_{l+1})
\label{cyclicity N}
\end{align}
form the cyclic OCHA $(\mathcal{H}_c\oplus\mathcal{H}_o,\Omega^o,\{L_n,N_{n,l}\})$.
Here, the sign factor $\mu_{m,i}(\sigma)$ in Eq.~(\ref{OCHA rel}) is given by
\begin{equation}
 \mu_{m,i}(\sigma)\ =\ \epsilon(\sigma) + 
\sum_{j=1}^{m}|\Phi_{\sigma(j)}| +
\sum_{j=1}^{i}|\Psi_{j}|(1 +
\sum_{k=m+1}^{n}|\Phi_{\sigma(k)}|).
\end{equation}
Note that the OCHA relation in Eq.~(\ref{OCHA rel}) includes the $A_\infty$ relation 
in Eq.~(\ref{A infinity}) as $n=0$, at which the cyclicity condition becomes the one 
for the open superstring in Eq.~(\ref{cyclicity open}).
In other words, the purely open-superstring interactions $N_{0,l}$ 
form the cyclic $A_\infty$ algebra $(\mathcal{H}_o,\Omega^o,\{N_{0,l}\})$.

The linear maps in Eq.~(\ref{linear map n}) are also represented by a degree odd coderivation
\begin{equation}
 \bd{N}=\sum_{\substack{n,l=0\\ n+l\ge1}}^\infty \bd{N}_{n,l}
\end{equation}
acting on $\mathcal{SH}_c^{res}\otimes\mathcal{TH}_o^{res}$ as
\begin{equation}
 \bd{N}\ =\ \sum_{\substack{n,l=0\\n+l>0}}\bd{N}_{n,l}\
=\ \sum_{\substack{n,l=0\\n+l>0}}\sum_{m,j,k=0}^\infty
\bigg(\id_m \otimes \Big(\id^{\otimes j}\otimes N_{n,l}\otimes \id^{\otimes k}\Big)\bigg)
\pi_{m+n,j+k+l},
\label{coderivation N}
\end{equation}
where $\pi_{n,l}$ is the projector onto the subspace 
$(\mathcal{H}_c^{res})^{\wedge n}\otimes (\mathcal{H}_o^{res})^{\otimes l}$.
By extending $\bd{L}$ to the coderivation acting on $\mathcal{SH}_c^{res}\otimes\mathcal{TH}_o^{res}$ as
\begin{equation}
 \bd{L}\ =\ \sum_{n=1}^\infty\ \bd{L}_n\
=\ \sum_{n=1}^\infty\sum_{m,l=0}^\infty 
\bigg(\Big(L_n\wedge\id_m\Big)\otimes \id^{\otimes l}\bigg)\pi_{n+m,l},
\end{equation}
we can consider the coderivation $\bd{L}+\bd{N}$. 
The OCHA relation in Eq.~(\ref{OCHA rel}) can then be written as
\begin{equation}
[\bd{L},\bd{N}]+\frac{1}{2}[\bd{N},\bd{N}]\ =\ 0,
\label{sOCHA rel}
\end{equation}
which we can rewrite as
\begin{equation}
 [\bd{L}+\bd{N},\bd{L}+\bd{N}]\ =\ 0,
\end{equation}
by combining with the $L_\infty$ relation $[\bd{L},\bd{L}]=0$. 
For open-closed superstring field theory, the string interaction $N_{n,l}$ must be defined
for any combination of four sectors of closed superstring and two sectors of open superstring
so that the sum of three kinds (open, left-moving, and right-moving) of picture numbers  are conserved:
\begin{equation}
\bd{N}\ =\ \sum_{\substack{p,n,m,r=0\\ n+m\ge1
}}^\infty  \bd{N}_{n;m}^{(p)}\mid_{2r}\delta_{p+r,2n+m-1},
\end{equation}
where $2r$ is the \textit{total} Ramond number defined by
\begin{align}
&\ \textrm{total Ramond number}\ =\
\nonumber\\
&\hspace{5mm}
 \textrm{\# of $\RNS$ inputs} +  \textrm{\# of $\NSR$ input} 
+ 2 (\textrm{\# of $\RR$ inputs})
+ \textrm{\# of open $R$ inputs} 
\nonumber\\
&\hspace{5mm} 
- \textrm{\# of open $R$ output}.
\nonumber
\end{align}

The action with OCHA structure is given by
\begin{equation}
 I_{oc}\ =\ \int^1_0 dt\, \Omega^o\left(\Psi,\pi_1
\bd{N}\left(e^{\wedge{\bf\Phi}}\otimes\frac{1}{1-t\Psi}\right)\right),
\label{action oc}
\end{equation}
which describes the open superstring field theory on the closed-superstring background
\footnote{
We omitted here the terms corresponding to a disk with closed strings in the bulk and no open strings 
on the boundary, which are included in the action proposed in Ref.~\cite{Zwiebach:1997fe}. These terms
give a constant determined by a closed string background but do not relevant to symmetry structure
of the theory \cite{Kajiura:2005sn}.}.
The open-superstring field $\Psi$ is dynamical and the closed-superstring field ${\bf\Phi}$ is
the background field satisfying the equation of motion
\begin{equation}
 \pi_1\bd{L}(e^{\wedge{\bf\Phi}})\ =\ 0.
\label{closed eom}
\end{equation}
The arbitrary variation of $I_{oc}$ is given by
\begin{equation}
 \delta I_{oc}\ =\ \Omega^o\left(\delta\Psi, \pi_1\bd{N}
\left(e^{\wedge{\bf\Phi}}\otimes
\frac{1}{1-\Psi}\right)\right).
\end{equation}
We can show that the action in Eq.~(\ref{action oc}) is invariant
under the gauge transformation
\begin{equation}
\delta_{\Lambda}\Psi\ =\ \pi_1\bd{N}\left(e^{\wedge{\bf \Phi}}\otimes
\left(\frac{1}{1-\Psi}\otimes
\Lambda\otimes\frac{1}{1-\Psi}\right)\right),
\label{gauge tf oc}
\end{equation}
using the relation in Eq.~(\ref{sOCHA rel}):
\begin{align}
 \delta_{\Lambda}I_{oc}\ =&\
\Omega^o\left(\Lambda,\pi_1\bd{N}\left(e^{\wedge{\bf\Phi}}\otimes\left(
\frac{1}{1-\Psi}\otimes\pi_1\bd{N}\left(e^{\wedge{\bf\Phi}}\otimes\frac{1}{1-\Psi}\right)
\otimes\frac{1}{1-\Psi}\right)\right)\right)
\nonumber\\
=&\ \Omega^o\left(\Lambda, \pi_1\bd{NN}\left(e^{\wedge{\bf\Phi}}\otimes\frac{1}{1-\Psi}\right)
\right)
\nonumber\\
=&\ -\Omega^o\left(\Lambda,\pi_1\bd{NL}\left(e^{\wedge{\bf\Phi}}\otimes\frac{1}{1-\Psi}\right)\right)
\nonumber\\
=&\ -\Omega^o\left(\Lambda, \pi_1\bd{N}\left(\left(e^{\wedge{\bf\Phi}}\wedge\pi_1\bd{L}\left(
e^{\wedge{\bf\Phi}}\right)\right)\otimes\frac{1}{1-\Psi}
\right)\right)\ =\ 0.
\end{align}
The open-closed superstring interaction $\bd{N}$ deforms 
by the background closed superstring field ${\bf\Phi}$ gives a weak $A_\infty$ algebra 
$(\mathcal{H}_o^{res},\bd{M}({\bf\Phi}))$ with
\begin{equation}
 \bd{M}({\bf\Phi})\ =\ \left(\bd{N}(e^{\bf\Phi}\otimes\id)\right)\pi^o.
\end{equation}
Here, $\id$ is the identity map in $\mathcal{TH}_o^{res}$
and $\pi^o$ is the projector onto $\mathcal{TH}_o^{res}$.

\subsection{Explicit construction of interactions}

Let us construct a cyclic OCHA 
$(\mathcal{H}_c^{res}\oplus\mathcal{H}_o^{res},\Omega^c\oplus\Omega^o,\bd{L}+\bd{N})$.
We assume that the cyclic sub-$L_\infty$-algebra
$\bd{L}$ is already constructed in the way given in the previous subsection.
Similarly to the previous cases,
we can construct $\bd{N}$ satisfying the relation in Eq.~(\ref{sOCHA rel}) and the cyclicity 
condition in Eq.~(\ref{cyclicity N}) in the following two steps. First consider a degree odd nilpotent
coderivation
\begin{equation}
\pi_1\bd{\mathcal{O}}\ 
=\ \pi_1\left(\bd{Q}-\bd{\eta}+\bd{A}+\bd{B}+\bd{C}\right)
-\left(1-\frac{1}{2}(X+\bar{X})\right)\pi_1^{(1,1)}\bd{B}
\label{OCHA O}
\end{equation}
satisfying $[\bd{\mathcal{O}},\bd{\mathcal{O}}]=0$,
or equivalently two mutually commutative coderivations
\begin{align}
 \pi_1\bd{\mathcal{D}}\ =&\ \pi_1\bd{Q}+\pi_1^{(0,0)}\bd{B}+\pi_1^0(\bd{A}+\bd{C}),\\
 \pi_1\bd{\mathcal{C}}\ =&\ \pi_1\bd{\eta}-\left(\pi_1^{(1,0)}+\pi_1^{(0,1)}+
\frac{1}{2}(X+\bar{X})\pi_1^{(1,1)}\right)\bd{B}-\pi_1^1(\bd{A}+\bd{C}),
\end{align}
satisfying $[\bd{\mathcal{D}},\bd{\mathcal{D}}]=[\bd{\mathcal{C}},\bd{\mathcal{C}}]
=[\bd{\mathcal{D}},\bd{\mathcal{C}}]=0$,
where $\bd{Q}$ acts as $\bd{Q}_c$ or $\bd{Q}_o$ on  $\mathcal{H}_c$ or $\mathcal{H}_o$,
respectively, and similarly $\bd{\eta}$ acts as  $\bd{\eta}+\bar{\bd{\eta}}$ or $\bd{\eta}$ on 
on  $\mathcal{H}_c$ or $\mathcal{H}_o$, respectively. 
Degree odd coderivations $\bd{A}$ and $\bd{B}$ are those for constructing $A_\infty$
and $L_\infty$ algebras in Eqs.~(\ref{A for A-inf}) and (\ref{B for L-inf}), and $\bd{C}$ is the one 
for constructing open-closed interaction
defined by respecting the cyclic Ramond number:
\begin{equation}
\bd{C}\ =\
\sum_{p,n,l,r=0}^\infty\delta_{p+r,2n+l+1}
\bd{C}_{n+1,l}^{(p)}\mid^{2r}.
\end{equation}
The OCHA relations can be written as 
the $L_\infty$ relations in Eq.~(\ref{L inf bare}) and the relations
\begin{subequations} \label{OCHA bare}
 \begin{align}
&\ [\bd{Q},\bd{C}] + [\bd{A},\bd{C}]^1
+ \frac{1}{2}[\bd{C},\bd{C}]^1 + [\bd{B},\bd{C}]^{11}\ =\ 0,\\
&\ [\bd{\eta},\bd{C}] - [\bd{A},\bd{C}]^2 - \frac{1}{2}[\bd{C},\bd{C}]^2 
- [\bd{B},\bd{C}]^{21} - [\bd{B},\bd{C}]^{12}
- [\bd{B},\bd{C}]^{22}_{X+\bar{X}}\ =\ 0.
 \end{align}
\end{subequations}
If we find such $\bd{A}$, $\bd{B}$, and $\bd{C}$, the cohomomorphism
\begin{align}
 \pi_1\hat{\bd{F}}^{-1}\ =&\
\pi_1\id - \left(\Xi^c\pi_1^{(1,0)} + \bar{\Xi}^c\pi_1^{(0,1)}
+\frac{1}{2}(\bar{\Xi}^cX^c+\Xi^c\bar{X}^c)\pi_1^{(1,1)}
\right)\bd{B} - \Xi^o\pi_1^1(\bd{A}+\bd{C})
\label{cohomo OCHA}
\end{align}
transforms $\bd{\mathcal{D}}$ and $\bd{\mathcal{C}}$ to the ones we eventually construct as
\begin{align}
 \pi_1\hat{\bd{F}}^{-1}\bd{\mathcal{D}}\hat{\bd{F}}\ =&\ \pi_1\bd{Q} + \mathcal{G}^c\pi_1\bd{B}\hat{\bd{F}}\
+ \mathcal{G}^0\pi_1(\bd{A}+\bd{C})\hat{\bd{F}}\
=\ \pi_1(\bd{L}+\bd{N}),\\
 \pi_1\hat{\bd{F}}^{-1}\bd{\mathcal{C}}\hat{\bd{F}}\ =&\ \bd{\eta}. 
\end{align}

We can construct $\bd{C}$ similarly to $\bd{A}$ and $\bd{B}$, which we already find.
By introducing parameters $s$ and $t$\,,
we consider a generating functions in Eq.~(\ref{gen fun A}), and Eq.~(\ref{gen fun L}),
and
\begin{align}
 \bd{C}(s,t)\ =&\ 
\sum_{p,m,n,l,r=0}^\infty
\delta_{m+p+r,2n+l+1}\,s^{m} t^{p} \bd{C}_{n+1,l}^{(p)}\mid^{2r}\
\equiv\
\sum_{p=0}^\infty t^p \bd{C}^{(p)}(s),
\end{align}
and extend the relations in Eqs.~(\ref{OCHA bare}) to
\begin{subequations} \label{ext OCHA rel}
 \begin{align}
 \bd{I}(s,t)\ \equiv&\ [\bd{Q},\bd{C}(s,t)] 
+ [\bd{A}(s,t),\bd{C}(s,t)]_{\mathfrak{o}_1(s)} 
+ \frac{1}{2}[\bd{C}(s,t),\bd{C}(s,t)]_{\mathfrak{o}_1(s)} 
\nonumber\\
&\hspace{62mm}
+ [\bd{B}(s,s,t),\bd{C}(s,t)]_{\mathfrak{c}_1(s,s,t)}\ =\ 0,\\
 \bd{J}(s,t)\ \equiv&\ [\bd{\eta},\bd{C}(s,t)] 
- [\bd{A}(s,t),\bd{C}(s,t)]_{\mathfrak{o}_2(t)}
- \frac{1}{2}[\bd{C}(s,t),\bd{C}(s,t)]_{\mathfrak{o}_2(t)}
\nonumber\\
&\hspace{61mm}
- [\bd{B}(s,s,t),\bd{C}(s,t)]_{\mathfrak{c}_2(s)+\bar{\mathfrak{c}}_2(s)}\ =\ 0.
\end{align}
\end{subequations}
We can show that
if
$\bd{C}(s,t)$ satisfy 
\begin{subequations}\label{diff OCHA} 
\begin{align}
 \partial_t\bd{C}(s,t)\ =&\ [\bd{Q},\bd{\nu}(s,t)]
\nonumber\\
&\
+ [\bd{A}(s,t),\bd{\nu}(s,t)]_{\mathfrak{o}_1(s)} 
+ [\bd{C}(s,t),\bd{\mu}(s,t)]_{\mathfrak{o}_1(s)}
+ [\bd{C}(s,t),\bd{\nu}(s,t)]_{\mathfrak{o}_1(s)}
\nonumber\\
&\
+ [\bd{B}(s,s,t),\bd{\nu}(s,t)]_{\mathfrak{c}_1(s,s,t)} 
+ [\bd{C}(s,t),(\bd{\lambda}+\bar{\bd{\lambda}})(s,s,t)]_{\mathfrak{c}_1(s,s,t)} 
\nonumber\\
&\
+ [\bd{B}(s,s,t),\bd{C}(s,t)]_{\mathfrak{d}(s,s)},\\
\partial_s\bd{C}(s,t)\ =&\ [\bd{\eta},\bd{\nu}(s,t)]
\nonumber\\
&\
- [\bd{A}(s,t),\bd{\nu}(s,t)]_{\mathfrak{o}_2(t)}
- [\bd{C}(s,t),\bd{\mu}(s,t)]_{\mathfrak{o}_2(t)}
- [\bd{C}(s,t),\bd{\nu}(s,t)]_{\mathfrak{o}_2(t)}
\nonumber\\
&\
- [\bd{B}(s,s,t),\bd{\nu}(s,t)]_{\mathfrak{c}_2(t)+\bar{\mathfrak{c}}_2(t)} 
- [\bd{C}(s,t),(\bd{\lambda}+\bar{\bd{\lambda}})(s,s,t)]_{\mathfrak{c}_2(t)+\bar{\mathfrak{c}}_2(t)} 
\end{align}
\end{subequations}
with degree even coderivation
\begin{equation}
 \bd{\nu}(s,t)\ =\ 
\sum_{p,n,l,r,m=0}^\infty\delta_{p+m+r,2n+l}\, s^m t^p \bd{\nu}_{n+1,l}^{(p+1)}\mid^{2r}\
\equiv\ \sum_{p=0}^\infty t^p \bd{\nu}^{(p+1)}(s),
\end{equation}
then, the $t$ derivative of the left hand sides of (\ref{ext OCHA rel}) become
\begin{subequations} 
 \begin{align}
\partial_t\bd{I}(s,t)\ =&\ [\bd{I}(s,t),(\bd{\mu}(s,t)+\bd{\nu}(s,t))]_{\mathfrak{o}_1(s)}
+ [\bd{I}(s,t),(\bd{\lambda}+\bar{\bd{\lambda}})(s,s,t)]_{\mathfrak{c}_1(s,s,t)}
\nonumber\\
&\
+ [\bd{I}(s,t), \bd{B}(s,s,t)]_{\mathfrak{d}(s,s)},\\
\partial_t\bd{J}(s,t)\ =&\ [\bd{J}(s,t),(\bd{\mu}(s,t)+\bd{\nu}(s,t))]_{\mathfrak{o}_1(s)}
+ [\bd{J}(s,t),(\bd{\lambda}+\bar{\bd{\lambda}})(s,s,t)]_{\mathfrak{c}_1(s,s,t)}
\nonumber\\
&\
+ [\bd{J}(s,t),\bd{B}\sst]_{\mathfrak{d}(s,s)} 
- \partial_s\bd{I}(s,t)
- [\bd{I}(s,t),(\bd{\mu}(s,t)+\bd{\nu}(s,t))]_{\mathfrak{o}_2(t)}\,,
 \end{align}
\end{subequations}
by using the differential equations (\ref{diff open}) and
\begin{align}
 \partial_t\bd{B}(s,s,t)\ =&\ [\bd{Q},(\bd{\lambda}+\bar{\bd{\lambda}})(s,s,t)]
\nonumber\\
&\
+ [\bd{B}(s,s,t),(\bd{\lambda}+\bar{\bd{\lambda}})(s,s,t)]_{\mathfrak{c}_1(s,s,t)}
+\frac{1}{2}[\bd{B}(s,s,t),\bd{B}(s,s,t)]_{\mathfrak{d}(s,s)}\,,\\
\partial_s\bd{B}(s,s,t)\ =&\ [\bd{\eta}+\bar{\bd{\eta}},(\bd{\lambda}+\bar{\bd{\lambda}})(s,s,t)]
- [\bd{B}(s,s,t),(\bd{\lambda}+\bar{\bd{\lambda}})(s,s,t)]_{\mathfrak{c}_2(t)+\bar{\mathfrak{c}}_2(t)},
\end{align}
satisfied by $\bd{A}(s,t)$ and $\bd{B}(s,s,t)$\,, respectively.
Therefore, if the relations at $t=0$
\begin{subequations} \label{OCHA boson}
 \begin{align}
 \bd{I}(s,0)\ =&\ [\bd{Q},\bd{C}(s,0)] 
+ [\bd{A}(s,0),\bd{C}(s,0)]_{\mathfrak{o}_1(s)}
+ \frac{1}{2}[\bd{C}(s,0),\bd{C}(s,0)]_{\mathfrak{o}_1(s)}
\nonumber\\
&\
+ [\bd{B}(s,s,0),\bd{C}(s,0)]_{\mathfrak{c}_1(s,s,0)}\ =\ 0,\\
 \bd{J}(s,0)\ =&\ [\bd{\eta},\bd{C}(s,0)]\ =\ 0, 
\end{align}
\end{subequations}
hold, then $\bd{I}(s,t)=\bd{J}(s,t)=0$ for any $t$. 
We can easily find that the coderivation $\bd{C}(s,0)$ satisfying Eq.~(\ref{OCHA boson})
has no picture number and is given by setting
\begin{equation}
 \bd{C}(s,0)\ =\ \bd{C}^{(0)}(s)\ =\ \bd{N}_B(s)\,,
\label{ini OCHA}
\end{equation}
with
\begin{equation}
 \bd{N}_B(s)\ =\ \sum_{m,n,l,r=0}^\infty 
s^{m}\delta_{m+r,2n+l+1}(\bd{N}_B)_{n+1,l}\mid^{2r},
\end{equation}
which can constructed similarly to those of the bosonic open-closed string field theory 
\cite{Zwiebach:1997fe}. Therefore, we can obtain $\bd{C}(s,t)$ satisfying Eq.~(\ref{ext OCHA rel})
by recursively solving the differential equations in Eqs.~(\ref{diff OCHA}) 
under the initial condition in Eq.~(\ref{ini OCHA}) to be cyclic with respect to $\omega_l^o$.
In Appendix~\ref{App A0}, we give a concrete procedure to solve them for some lower orders.
The cyclic OCHA $(\mathcal{H}^{res},\Omega,\bd{L}+\bd{N})$ is eventually constructed by
transforming using cohomomorphism in Eq.~(\ref{cohomo OCHA}).

\sectiono{Mapping to WZW-like action}\label{D}

The WZW-like formulation is the other complementary way to construct superstring field theories
using the large Hilbert space \cite{Berkovits:1995ab,Erler:2015rra,Kunitomo:2015usa,Berkovits:2004xh,Matsunaga:2016zsu,
Erler:2017onq,Kunitomo:2019glq,Kunitomo:2019kwk,Kunitomo:2021wiz}.
We can map 
the action we constructed in the previous section to the WZW-like action as in the open, heterotic, and 
type II superstring field theories
\cite{Erler:2015rra,Kunitomo:2019glq,Kunitomo:2019kwk,Kunitomo:2021wiz}.

Let us first focus on the NS $\oplus$ NS-NS sector, which we simply call the NS sector in this section.
The map between two formulations, the homotopy-based and WZW-like formulations, is given 
by the cohomomorphism $\hat{\bd{g}}=\hat{\bd{g}}_c\otimes\hat{\bd{g}}_o$ 
\cite{Erler:2016ybs,Kunitomo:2019glq,Kunitomo:2021wiz} with
\begin{equation}
\hat{\bd{g}}_c\ =\ \vec{\mathcal{P}}\exp\left(
\int^1_0dt\,(\bd{\lambda}+\bar{\bd{\lambda}})^{NS}(0,t)\right),\qquad
\hat{\bd{g}}_o\ =\ \vec{\mathcal{P}}\exp\left(
\int^1_0dt\,\bd{\mu}^{NS}(0,t)\right),
\end{equation}
where
\begin{align}
 (\bd{\lambda}+ \bar{\bd{\lambda}})^{NS}(s,t)\ =&\ 
\sum_{m,p=0}^\infty\sum_{\bar{m},\bar{p}=0}^\infty
s^{m+\bar{m}}t^{p+\bar{p}} \left(
\bd{\lambda}^{(p+1,\bar{p})}_{m+p+2,\bar{m}+\bar{p}+1}\mid^{(0,0)}
+ \bar{\bd{\lambda}}^{(p,\bar{p}+1)}_{m+p+1,\bar{m}+\bar{p}+2}\mid^{(0,0)}\right),\\
 \bd{\mu}^{NS}(s,t)\ =&\ \sum_{m,p=0}^\infty s^m t^p \bd{\mu}_{m+p+2}^{(p+1)}\mid^{0}.
\end{align}
This cohomomorphism maps the string fields $(\Phi_{\NSNS},\Psi_{NS})$ to those 
in the WZW-like formulation $(V_o,V_c)$ as
\begin{equation}
 \pi_1^c\hat{\bd{g}}_c(e^{\wedge\Phi_{\NSNS}})\ =\ G_c(V_c),\qquad
 \pi_1^o\hat{\bd{g}}_o\left(\frac{1}{1-\Psi_{NS}}\right)\ =\ G_o(V_o),
\end{equation}
where 
\begin{equation}
G_c(V)\ =\ 
\eta\bar{\eta} V + \frac{1}{2}\left(
L^\eta_2(\eta\bar{\eta} V,\bar{\eta}V)+\eta L_2^{\bar{\eta}}(\eta\bar{\eta}V,V)
\right) + \cdots,
\end{equation}
is the pure-gauge string fields of type II superstring identically satisfying
\begin{equation}
\bd{L}_c^\eta(e^{\wedge G_c(V_c)})\ =\ 0,\qquad
\bd{L}_c^{\bar{\eta}}(e^{\wedge G_c(V_c)})\ =\ 0,
\end{equation}
with $(\bd{L}_c^\eta,\bd{L}_c^{\bar{\eta}})=(\hat{\bd{g}}_c\bd{\eta}_c\hat{\bd{g}}_c^{-1},
\hat{\bd{g}}_c\bar{\bd{\eta}}_c\hat{\bd{g}}_c^{-1})$ \cite{Matsunaga:2016zsu}.
The pure-gauge string field $G_o(V_o)$ of the open superstring is similarly defined by
a composite string field of $V_o$ identically satisfying the equation
\begin{equation}
 \bd{L}_o^\eta\left(\frac{1}{1-G_o(V_o)}\right)\ =\ 0,
\end{equation}
with $\bd{L}_o^\eta=\hat{\bd{g}}_o\bd{\eta}_o\hat{\bd{g}}_o^{-1}$.
We give a prescription to obtain explicit form of 
$G_o(V_o)$ in Appendix~\ref{App A}.
The (dynamical) equation of motion 
of the open superstring 
is mapped as
\begin{equation}
 \pi_1\tilde{\bd{N}}_{NS}\left(e^{\wedge G_c(V_c)}\otimes\frac{1}{1-G_o(V_o)}\right)\ =\ 0,
\label{eom wzw}
\end{equation}
with
\begin{equation}
\tilde{\bd{N}}_{NS}\ =\ \hat{\bd{g}}\bd{N}_{NS}\hat{\bd{g}}^{-1}\ =\ 
\bd{Q}_o+\hat{\bd{g}}(\bd{N}_{NS}-\bd{M}_{NS})\hat{\bd{g}}^{-1},   
\end{equation}
where 
$V_c$
is a background field satisfying the equation of motion of the closed-superstring $Q_cG_c(V_c)=0$. 
In order to give the WZW-like action deriving this equation of motion in Eq.~(\ref{eom wzw}), we define the associated
string field as
\begin{equation}
 B_d(V_o)\ =\ \pi_1^o\hat{\bd{g}}_o\bd{\xi_d}\left(\frac{1}{1-\Psi_{NS}}\right),
\end{equation}
where $d=t, \delta$ or $Q$ and $\bd{\xi_d}$ is the coderivation derived from 
$\xi\partial_t$, $\xi\delta$ or $-\xi\pi_1\bd{M}_{NS}$, respectively. We can show that the relations
\begin{align}
&\  dG_o(V_o)\ =\ (-1)^d D_\eta B_d(V_o),\label{dG}\\
&\ D_\eta\left(\partial_tB_\delta(V_o)-\delta B_{\partial_t}(V_o)\right)\ =\ 0.
\end{align}
hold\footnote{Note that deg$(V_o)=1$ and deg$(B_d)=$deg$(d)+1$.}, where 
$D_{\eta}$ is the nilpotent linear operator defined by
\begin{align}
 D_\eta\varphi\ 
=&\ \pi_1^o\bd{L}_o^\eta\left(\frac{1}{1-G_o(V_o)}\otimes\varphi\otimes\frac{1}{1-G_o(V_o)}\right)
\nonumber\\
=&\ \pi_1^o\bd{L}^\eta\left(e^{\wedge G_c(V_c)}\otimes
\left(
\frac{1}{1-G_o(V_o)}\otimes\varphi\otimes\frac{1}{1-G_o(V_o)}\right)\right),
\end{align}
acting on an open superstring field $\varphi\in\mathcal{H}_{NS}$. The coderivation
$\bd{L}^{\eta}$ acts as $\bd{L}_c^{\eta}+\bd{L}_c^{\bar{\eta}}$ on $\mathcal{H}_c$ and as $\bd{L}_o^{\eta}$ on
$\mathcal{H}_o$.
Then, the WZW-like action for the NS sector is given by
\begin{equation}
 I_{WZW}^{NS}\ =\ \int_0^1 dt\, \omega_l^o\left(B_{t}(t V_o), 
\pi_1\tilde{\bd{N}}_{NS}\Big(e^{\wedge G_c(V_c)}\otimes\frac{1}{1-G_o(t V_o)}\Big)
\right),
\end{equation}
which is invariant under the gauge transformation
\begin{align}
 B_\delta(V_o)\ =&\ \pi_1\tilde{\bd{N}}_{NS}\left(e^{\wedge G_c(V_c)}\otimes
\left(
\frac{1}{1-G_o(V_o)}\otimes\Lambda\otimes\frac{1}{1-G_o(V_o)}\right)\right) 
+ D_\eta\Omega.
\end{align}

It is straightforward to extend these results of the NS sector to all the sectors.
Since $\hat{\bd{g}}_c$ and $\hat{\bd{g}}_o$ act as the identity operators outside the NS sector, we find that
\begin{align}
 \pi_1^c\hat{\bd{g}}_c\left(e^{\wedge \Phi}\right)\ =&\ \pi_1^c\hat{\bd{g}}_c\left(e^{\wedge \Phi_{\NSNS}}\right) 
+ \Phi_{\RNS} + \Phi_{\NSR} + \Phi_{\RR},\\
 \pi_1^o\hat{\bd{g}}_o\left(\frac{1}{1-\Psi}\right)\ =&\ 
\pi_1^o\hat{\bd{g}}_o\left(\frac{1}{1-\Psi_{NS}}\right) + \Psi_{R},
\end{align}
and can identify the components $(\Phi_{\RNS},\Phi_{\NSR},\Phi_{\RR};\Psi_R)$ to those in the WZW-like formulation
$(\Psi_c, \bar{\Psi}_c,\Sigma_c;\Psi_o)$:
\begin{equation}
\Phi_{\RNS}\ =\ \Psi_c,\qquad \Phi_{\NSR}\ =\ \bar{\Psi}_c,\qquad 
\Phi_{\RR}\ =\ \Sigma_c,\qquad \Psi_R\ =\ \Psi_o.
\end{equation}
Thus, these components are also annihilated by $\eta_c$ and $\bar{\eta}_c$ (or $\eta_o$)
and satisfy the constraint in Eq.~(\ref{closed constraint}) 
(or Eq.~(\ref{open constraint})).
%
The WZW-like action of the open superstring field theory on the general closed-string backgrounds 
is eventually written as
\begin{align}
 I_{WZW}\ =&\ \int^1_0 dt\,\omega_l^o\Bigg(\mathcal{B}_{t}(\mathcal{V}_o(t)),
(\mathcal{G}^o)^{-1}\pi_1\tilde{\bd{N}}\bigg(e^{\wedge (G(V_c)+\Psi_c+\bar{\Psi}_c+\Sigma_c)}\otimes
\frac{1}{1-G_o(V_o(t))-\Psi_o(t)}\bigg)\Bigg),
\label{WZW}
\end{align}
where $\tilde{\bd{N}}=\hat{\bd{g}}\bd{N}\hat{\bd{g}}^{-1}$ and
\begin{equation}
 \mathcal{B}_{t}(\mathcal{V}_o(t))\ =\ B_{t}(V_o(t))+\xi_0\partial_t\Psi_o(t).
\end{equation}
The closed superstring backgrounds $(V_c,\Psi_c,\bar{\Psi}_c,\Sigma_c)$ satisfy
\begin{equation}
 \pi_1\tilde{\bd{L}}\left(e^{\wedge(G_c(V_c)+\Psi_c+\bar{\Psi}_c+\Sigma_c)}\right)\ =\ 0
\end{equation}
with $\tilde{\bd{L}}=\hat{\bd{g}}\bd{L}\hat{\bd{g}}^{-1}$.
Note that, since $\hat{\bd{g}}$ acts as the identity except on the NS sector,
$\tilde{\bd{N}}$ and $\tilde{\bd{L}}$ preserve the constraints in Eqs.~(\ref{open constraint})
and (\ref{closed constraint}), respectively.
%
The WZW-like action in Eq.~(\ref{WZW}) is invariant under the gauge transformation
\begin{align}
  \mathcal{B}_\delta(\mathcal{V}_o)\ =&\ \pi_1\tilde{\bd{N}}
\left(e^{\wedge(G_c(V_c)+\Psi_c+\bar{\Psi}_c+\Sigma_c)}\otimes
\left(
\frac{1}{1-G_o(V_o)-\Psi_o}\otimes
(\Lambda+\xi\lambda)\otimes
\frac{1}{1-G_o(V_o)-\Psi_o}\right)\right),
\end{align}
which is also obtained through the map $\hat{\bd{g}}$. 
Here, $\Lambda$ and $\lambda$ are the gauge parameters in the NS and R sectors, respectively,
and $\lambda$ is annihilated by $\eta$ and satisfies the constraint in Eq.~(\ref{open constraint}).

\sectiono{Summary and discussion}\label{E}

In this paper, we constructed interactions for the open-closed superstring 
field theory based on the OCHA structure. It provides the open-closed superstring 
field theory on general closed-superstring backgrounds. 
We also give a corresponding WZW-like action for open-closed superstring field
theory through a field redefinition.

Recently, the open string field theory deformed with a gauge invariant open-closed coupling
is studied \cite{Erbin:2020eyc,Koyama:2020qfb,Maccaferri:2021ulf,Maccaferri:2021ksp,Maccaferri:2021lau}.
The effective open superstring field theory is governed by a weak $A_\infty$ structure
which includes non-trivial tadpole term, destabilizing the initial perturbative vacuum.
It requires to shift the vacuum to a new equilibrium point.
The open-closed superstring field theory, given in this paper, provides a basis for such 
an analysis on more general closed-superstring backgrounds described by classical solutions 
of the nonlinear equation of motion of the closed superstring field theory. 

In order to quantize the classical superstring field theory, we must
extend the classical action to the quantum master action satisfying the quantum BV equation. 
Such an open-closed superstring field theory is recently given in Ref.~\cite{FarooghMoosavian:2019yke}
based on the formalism using the extra free field \cite{Sen:2015hha,Sen:2015uaa}.
It is interesting to give a quantum master action using the formulation based on the homotopy
algebra, which requires to extend the OCHA structure to the quantum OCHA structure 
\cite{Munster:2011ij}.
The quantum open-closed superstring field theory is also practically useful
to study the string dynamics on the Ramond-Ramond backgrounds \cite{Cho:2018nfn},
the D-brane backgrounds \cite{Sen:2019qqg,Sen:2020cef,Sen:2020ruy,Sen:2020oqr,Sen:2021qdk}, 
and  so on, which are difficult in the first-quantized formulation using the RNS formalism.
The (quantum) OCHA structure should shed new light on such nonperturbative studies.

\vspace{1cm}
\noindent
{\bf \large Acknowledgments}

The author would like to thank Jojiro Yoshinaka for pointing out an error in the draft.
This work is supported in part by JSPS Grant-in-Aid for Scientific 
Research (C) Grant Number JP18K03645.

\vspace{1cm}
\appendix

\sectiono{Explicit procedure for solving
(\ref{diff closed}) and (\ref{diff OCHA})
}\label{App A0}

Similar to the open superstring case in section~\ref{A}, the differential equations 
(\ref{diff closed}) for constructing a $L_\infty$ algebra can be solved recursively 
with the initial condition in Eq.~(\ref{closed ini}).
The concrete procedure is, however, complicated due to the fact that
the parameter $t$ counts only the total picture number without independently
counting the left- and right-moving picture numbers.
We first rewrite the differential equations (\ref{diff closed}) in the form
\begin{align}
&\hspace{-20mm}
 \sum_{q=0}^{p+1}(p+1)\bd{B}^{(p-q+1,q)}\ssb\ 
\nonumber\\
=&\ 
\sum_{q=0}^p[\bd{Q},(\bd{\lambda}^{(p-q+1,q)}+\bar{\bd{\lambda}}^{(p-q,q+1)})\ssb]
\nonumber\\
&\ 
+ \sum_{q=0}^p\sum_{l=0}^q\sum_{m=0}^l
[\bd{B}^{(p-q,q-l)}\ssb,(\bd{\lambda}^{(l-m+1,m)}+\bar{\bd{\lambda}}^{(l-m,m+1)})\ssb]_{\mathfrak{c}_1^0\ssb}
\nonumber\\
&\
+ \sum_{q=0}^{p-1}\sum_{l=0}^q\sum_{m=0}^l
[\bd{B}^{(p-q-1,q-l)}\ssb,(\bd{\lambda}^{(l-m+1,m)}+\bar{\bd{\lambda}}^{(l-m,m+1)})\ssb]_{\mathfrak{c}_1^1\ssb}
\nonumber\\
&\
+ \frac{1}{2}\sum_{q=0}^p\sum_{l=0}^q\sum_{m=0}^l
[\bd{B}^{(p-q,q-l)}\ssb,\bd{B}^{(l-m,m)}\ssb]_{\mathfrak{d}\ssb}\,,
\label{eq1}\\
&\hspace{-20mm}
\sum_{q=0}^p[\bd{\eta},\bd{\lambda}^{(p-q+1,q)}\ssb]
\nonumber\\
=&\
\sum_{q=0}^p\partial_s\bd{B}^{(p-q,q)}\ssb\ 
\nonumber\\
&\
-\sum_{q=0}^{p-1}\sum_{l=0}^q\sum_{m=0}^l
[\bd{B}^{(p-q-1,q-l)}\ssb,(\bd{\lambda}^{(l-m+1,m)}+\bar{\bd{\lambda}}^{(l-m,m+1)})\ssb]^{21}
\nonumber\\
&\
-\sum_{q=0}^{p-2}\sum_{l=0}^k\sum_{m=0}^l
[\bd{B}^{(p-q-2,q-l)}\ssb,(\bd{\lambda}^{(l-m+1,m)}+\bar{\bd{\lambda}}^{(l-m,m+1)})\ssb]^{22}_{\bar{X}}\,,
\label{eq2}\\
&\hspace{-20mm}
\sum_{q=0}^p[\bar{\bd{\eta}},\bar{\bd{\lambda}}^{(p-q,q+1)}\ssb]
\nonumber\\
=&\ 
 \sum_{q=0}^p\partial_{\bar{s}}\bd{B}^{(p-q,q)}\ssb\ 
\nonumber\\
&\
-\sum_{q=0}^{p-1}\sum_{l=0}^q\sum_{m=0}^l
[\bd{B}^{(p-q-1,q-l)}\ssb,(\bd{\lambda}^{(l-m+1,m)}+\bar{\bd{\lambda}}^{(l-m,m+1)})\ssb]^{12}
\nonumber\\
&\
-\sum_{q=0}^{p-2}\sum_{l=0}^q\sum_{m=0}^l
[\bd{B}^{(p-q-2,q-l)}\ssb,(\bd{\lambda}^{(l-m+1,m)}+\bar{\bd{\lambda}}^{(l-m,m+1)})\ssb]^{22}_{X}\,,
\label{eq3}
\end{align}
where we expanded $\mathfrak{c}_1\sst$ in the power of $t$ as 
$\mathfrak{c}_1\sst=\mathfrak{c}_1^0\ssb+t\mathfrak{c}_1^1\ssb$ with
$\mathfrak{c}_1^0(s,\bar{s})=\pi^{(0,0)}+s\pi^{(1,0)}+ \bar{s}\pi^{(0,1)}+s\bar{s}\pi^{(1,1)}$
and $\mathfrak{c}_1^1\ssb=(s\bar{X}+\bar{s}X)\pi^{(1,1)}$\,.
The first one (\ref{eq1}) determines several 
$\bd{B}^{(p,\bar{p})}(s,\bar{s})$ with the same total picture number simultaneously.
We must split them by each left- and right-moving picture number. The explicit decomposition for 
the NS-NS sector was given in Ref.~\cite{Erler:2014eba}, but we have not yet extend it
to the whole sectors in a closed form. Instead, we give an explicit decomposition for some 
lower picture numbers and show how the equations determine $\bd{B}^{(p,\bar{p})}\ssb$ 
for all the higher picture numbers.
First, setting $p=0$ in Eqs.~(\ref{eq2}) and (\ref{eq3}), we have
\begin{equation}
[\bd{\eta},\bd{\lambda}^{(1,0)}\ssb]\ =\ \partial_s\bd{B}^{(0,0)}\ssb\,,\qquad 
[\bar{\bd{\eta}},\bar{\bd{\lambda}}^{(0,1)}\ssb]\ =\ \partial_{\bar{s}}\bd{B}^{(0,0)}\ssb\,.
\end{equation} 
We can solve them as
\begin{equation}
 \bd{\lambda}^{(1,0)}\ssb\ =\
\xi_0\circ\partial_s\bd{L}_B\ssb\,,\qquad
\bar{\bd{\lambda}}^{(0,1)}\ssb\ =\
\bar{\xi}_0\circ\partial_{\bar{s}}\bd{L}_B\ssb\,,
\label{lambda1}
\end{equation}
under the initial conditions $\bd{B}^{(0,0)}\ssb=\bd{L}_B\ssb$\,.
The operations $\xi_0^c\circ$ and $\bar{\xi}_0^c\circ$ are defined on general coderivation 
$\bd{B}=\sum_{n=0}^\infty\bd{B}_{n+2}$ by
\begin{subequations}\label{ximaru} 
\begin{align}
 \xi_0^c\circ\bd{B}\ =&\ \sum_{n,k=0}^\infty\Big(
(\xi_0^c\circ B_{n+2})\wedge \id_k\Big)\pi^c_{n+k+2}\,,\\
\xi_0^c\circ B_{n+2}\ =&\ \frac{1}{n+3}\Big(
\xi_0^c B_{n+2}-(-1)^{\textrm{deg}(B)}B_{n+2}(\xi_0^c\wedge\id_{n+1})
\Big)\,,
\end{align}
\end{subequations}
and those replacing
$\xi_0^c$ with $\bar{\xi}_0^c$.
Eq.~(\ref{eq1}) at $p=0$ splits into two equations
\begin{align}
 \bd{B}^{(1,0)}\ssb =&\ [\bd{Q},\bd{\lambda}^{(1,0)}\ssb]
+ [\bd{L}_B\ssb,\bd{\lambda}^{(1,0)}\ssb]_{\mathfrak{c}_1^0\ssb}
+\frac{\bar{s}}{2}[ 
\bd{L}_B\ssb,\bd{L}_B\ssb]^{22}_{\Xi}\,,\\
 \bd{B}^{(0,1)}\ssb =&\ [\bd{Q},\bar{\bd{\lambda}}^{(0,1)}\ssb]
+ [\bd{L}_B\ssb,\bar{\bd{\lambda}}^{(0,1)}\ssb]_{\mathfrak{c}_1^0\ssb}
+\frac{s}{2}[ 
\bd{L}_B\ssb,\bd{L}_B\ssb]^{22}_{\bar{\Xi}}\,.
\end{align}
Substituting Eq.~(\ref{lambda1}) in these expression, we obtain
$\bd{B}^{(1,0)}\ssb$ and $\bd{B}^{(0,1)}\ssb$ independently.
Next, setting $p=1$ in Eqs.~(\ref{eq2}) and (\ref{eq3}), we have
\begin{align}
 [\bd{\eta},\bd{\lambda}^{(2,0)}\ssb]\ =&\
\partial_s\bd{B}^{(1,0)}\ssb 
+ [\bd{L}_B\ssb,\bd{\lambda}^{(1,0)}\ssb]^{21}\,,\\
 [\bd{\eta},\bd{\lambda}^{(1,1)}\ssb]\ =&\
\partial_s\bd{B}^{(0,1)}\ssb
+ [\bd{L}_B\ssb,\bar{\bd{\lambda}}^{(0,1)}\ssb]^{21}\,,\\
 [\bar{\bd{\eta}},\bar{\bd{\lambda}}^{(1,1)}\ssb]\ =&\
\partial_{\bar{s}}\bd{B}^{(1,0)}\ssb
+ [\bd{L}_B\ssb,\bd{\lambda}^{(1,0)}\ssb]^{12}\,,\\
 [\bar{\bd{\eta}},\bar{\bd{\lambda}}^{(0,2)}\ssb]\ =&\
\partial_{\bar{s}}\bd{B}^{(0,1)}\ssb
+ [\bd{L}_B\ssb,\bar{\bd{\lambda}}^{(0,1)}\ssb]^{12}\,.
\end{align}
Since all the quantities in the right hand sides are already known, 
we can solve these equations for
$\bd{\lambda}^{(2,0)}\ssb$\,, $\bd{\lambda}^{(1,1)}\ssb$\,, 
$\bar{\bd{\lambda}}^{(1,1)}\ssb$\,, and $\bar{\bd{\lambda}}^{(0,2)}\ssb$ 
by acting $\xi_0^c\circ$ or $\bar{\xi}_0^c\circ$ as
\begin{align}
 \bd{\lambda}^{(2,0)}\ssb\ =\
\xi_0^c\circ\Big(
\partial_s\bd{B}^{(1,0)}\ssb 
+ [\bd{L}_B\ssb,\bd{\lambda}^{(1,0)}\ssb]^{21}
\Big)\,,\\
 \bd{\lambda}^{(1,1)}\ssb\ =\
\xi_0^c\circ\Big(
\partial_s\bd{B}^{(0,1)}\ssb 
+ [\bd{L}_B\ssb,\bd{\lambda}^{(0,1)}\ssb]^{21}
\Big)\,,\\
 \bar{\bd{\lambda}}^{(1,1)}\ssb\ =\
\bar{\xi}_0^c\circ\Big(
\partial_{\bar{s}}\bd{B}^{(1,0)}\ssb 
+ [\bd{L}_B\ssb,\bd{\lambda}^{(1,0)}\ssb]^{12}
\Big)\,,\\
 \bar{\bd{\lambda}}^{(0,2)}\ssb\ =\
\bar{\xi}_0^c\circ\Big(
\partial_{\bar{s}}\bd{B}^{(0,1)}\ssb 
+ [\bd{L}_B\ssb,\bd{\lambda}^{(0,1)}\ssb]^{12}
\Big)\,.
\end{align}
At $p=1$, Eq.~(\ref{eq1}) can be split as
\begin{subequations} \label{Bn1}
\begin{align}
 2\bd{B}^{(2,0)}\ssb\ =&\ 
[\bd{Q},\bd{\lambda}^{(2,0)}\ssb]
+[\bd{B}^{(1,0)}\ssb,\bd{\lambda}^{(1,0)}\ssb]_{\mathfrak{c}_1^0\ssb}
\nonumber\\
&\
+[\bd{B}^{(0,0)}\ssb,\bd{\lambda}^{(2,0)}\ssb]_{\mathfrak{c}_1^0\ssb}
+\bar{s} [\bd{B}^{(0,0)}\ssb,\bd{\lambda}^{(1,0)}\ssb]^{22}_{X}
\nonumber\\
&\
+ \bar{s} [\bd{B}^{(0,0)}\ssb,\bd{B}^{(1,0)}\ssb]^{22}_{\Xi}\,,\\
 2\bd{B}^{(1,1)}\ssb\ =&\
[\bd{Q},(\bd{\lambda}^{(1,1)}+\bar{\bd{\lambda}}^{(1,1)})\ssb]
+[\bd{B}^{(0,0)}\ssb,(\bd{\lambda}^{(1,1)}+\bar{\bd{\lambda}}^{(1,1)})\ssb]_{\mathfrak{c}_1^0\ssb}
\nonumber\\
&\
+[\bd{B}^{(0,1)}\ssb,\bd{\lambda}^{(1,0)}\ssb]_{\mathfrak{c}_1^0\ssb}
+[\bd{B}^{(1,0)}\ssb,\bar{\bd{\lambda}}^{(0,1)}\ssb]_{\mathfrak{c}_1^0\ssb}
\nonumber\\
&\
+ s [\bd{B}^{(0,0)}\ssb,\bd{\lambda}^{(1,0)}\ssb]^{22}_{\bar{X}}
+ \bar{s} [\bd{B}^{(0,0)}\ssb,\bar{\bd{\lambda}}^{(0,1)}\ssb]^{22}_{X}
\nonumber\\
&\
+ s [\bd{B}^{(0,0)}\ssb,\bd{B}^{(1,0)}\ssb]^{22}_{\bar{\Xi}}
+ \bar{s} [\bd{B}^{(0,0)}\ssb,\bd{B}^{(0,1)}\ssb]^{22}_{\Xi}\,,
\\
 2\bd{B}^{(0,2)}\ssb\ =&\
[\bd{Q},\bar{\bd{\lambda}}^{(0,2)}\ssb]
+[\bd{B}^{(0,1)}\ssb,\bar{\bd{\lambda}}^{(0,1)}\ssb]_{\mathfrak{c}_1^0\ssb}
\nonumber\\
&\
+[\bd{B}^{(0,0)}\ssb,\bar{\bd{\lambda}}^{(0,2)}\ssb]_{\mathfrak{c}_1^0\ssb}
+ s [\bd{B}^{(0,0)}\ssb,\bar{\bd{\lambda}}^{(0,1)}\ssb]^{22}_{\bar{X}}
\nonumber\\
&\
+ s [\bd{B}^{(0,0)}\ssb,\bd{B}^{(0,1)}\ssb]^{22}_{\bar{\Xi}}\,.
\end{align}
\end{subequations}
All the quantities in the right hand sides have been obtained in 
the previous steps, and thus, Eqs.~(\ref{Bn1})
determine $\bd{B}^{(2,0)}\ssb$\,, $\bd{B}^{(2,0)}\ssb$\,, 
and $\bd{B}^{(2,0)}\ssb$\,.
Repeating the procedure, we can obtain
$\bd{B}^{(p,\bar{p})}\ssb$ for arbitrary $p$ and $\bar{p}$ independently.
Similar but slightly different analysis was give in Ref.~\cite{Kunitomo:2021wiz}.

The differential equations in Eqs.~(\ref{diff OCHA}) for open-closed interactions 
is also solved recursively with the initial condition in Eq.~(\ref{ini OCHA}).
The differential equations in Eqs.~(\ref{diff OCHA}) are rewritten as
\begin{align}
 (p+1)\bd{C}^{(p+1)}(s)\
=&\ [\bd{Q},\bd{\nu}^{(p+1)}(s)]
\nonumber\\
&\hspace{-15mm}
+\sum_{q=0}^p\Big(
[\left(\bd{A}^{(p-q)}(s)+\bd{C}^{(p-q)}(s)\right),\bd{\nu}^{(q+1)}(s)]_{\mathfrak{o}_1(s)}
+[\bd{C}^{(p-q)}(s),\bd{\mu}^{(q+1)}(s)]_{\mathfrak{o}_1(s)}
\Big)
\nonumber\\
&\hspace{-15mm}
+\sum_{q=0}^p\sum_{l=0}^{p-q}\Big(
[\bd{B}^{(p-q-l,q)}(s,s),\bd{\nu}^{(l+1)}(s)]_{\mathfrak{c}_1^0(s,s)}
\nonumber\\
&\hspace{3cm}
+ [\bd{C}^{(p-q-l)}(s),
(\bd{\lambda}^{(l+1,q)}+\bar{\bd{\lambda}}^{(q,l+1)})(s,s)]_{\mathfrak{c}_1^0(s,s)}
\Big)
\nonumber\\
&\hspace{-15mm}
+\sum_{q=0}^{p-1}\sum_{l=0}^{p-q-1}s \Big(
[\bd{B}^{(p-q-l-1,q)}(s,s),\bd{\nu}^{(l+1)}(s)]^{22}_{X+\bar{X}}
\nonumber\\
&\hspace{3cm}
+ [\bd{C}^{(p-q-l-1)}(s),
(\bd{\lambda}^{(l+1,q)}+\bar{\bd{\lambda}}^{(q,l+1)})(s,s)]^{22}_{X+\bar{X}}
\Big)
\nonumber\\
&\hspace{-15mm}
+ \sum_{q=0}^p\sum_{l=0}^{p-q} s [\bd{B}^{(p-q-l,q)}(s,s),
\bd{C}^{(l)}(s)]^{22}_{\Xi+\bar{\Xi}}\,,
\label{app ocha 1}
\end{align}
\begin{align}
[\bd{\eta},\bd{\nu}^{(p+1)}(s)]\
=&\
\partial_s\bd{C}^{(p)}(s) 
\nonumber\\
&\hspace{-15mm}
+ \sum_{q=0}^{p-1}\Big(
[\left(\bd{A}^{(p-q-1)}(s)+\bd{C}^{(p-q-1)}(s)\right),\bd{\nu}^{(q+1)}(s)]^2
+ [\bd{C}^{(p-q-1)}(s),\bd{\mu}^{(q+1)}(s)]^2
\Big)
\nonumber\\
&\hspace{-15mm}
+ \sum_{q=0}^{p-1}\sum_{l=0}^{p-q-1}\Big(
[\bd{B}^{(p-q-l-1,q)}(s,s),\bd{\nu}^{(l+1)}(s)]^{21+12}
\nonumber\\
&\hspace{3cm}
+ [\bd{C}^{(p-q-l-1)}(s),(\bd{\lambda}^{(l+1,q)}+\bar{\bd{\lambda}}^{(q,l+1)})(s,s)]^{21+12}
\Big)
\nonumber\\
&\hspace{-15mm}
+ \frac{1}{2}\sum_{q=0}^{p-2}\sum_{l=0}^{p-q-2}\Big(
[\bd{B}^{(p-q-l-2,q)}(s,s),\bd{\nu}^{(l+1)}(s)]^{22}_{X+\bar{X}}
\nonumber\\
&\hspace{3cm}
+ [\bd{C}^{(p-q-l-2)}(s),(\bd{\lambda}^{(l+1,q)}+\bar{\bd{\lambda}}^{(q,l+1)})(s,s)]^{22}_{X+\bar{X}}
\Big)\,,
\label{app ocha 2}
\end{align}
where  we denote
$[A,B]^{21}+[A,B]^{12}$ as $[A,B]^{21+12}$ for notational simplicity\,.
We assume that $\bd{A}^{(p)}(s)$\,, $\bd{B}^{(p,\bar{p})}(s,\bar{s})$\,, $\bd{\mu}^{(p+1)}(s)$\,,
$\bd{\lambda}^{(p+1,\bar{p})}(s,\bar{s})$\,, and $\bar{\bd{\lambda}}^{(p,\bar{p}+1)}(s,\bar{s})$ 
are independently determined by
solving the differential equations in Eqs.~(\ref{diff open}) and (\ref{diff closed}).

We start from Eq.~(\ref{app ocha 2}) at $p=0$ with the initial condition in Eq.~(\ref{ini OCHA}):
\begin{equation}
 [\bd{\eta},\bd{\nu}^{(1)}(s)]\ =\ \partial_s\bd{N}_B(s)\,.
\end{equation}
This is solved as
\begin{equation}
 \bd{\nu}^{(1)}(s)\ =\ \xi_0^o\circ\partial_s\bd{N}_B(s)\,,
\label{nu 1}
\end{equation}
so as to respect the cyclicity,
where $\xi_0^o\circ$ is defined on general coderivation 
$\bd{C}=\sum_{n,l=0}^\infty\bd{C}_{n+1,l}$ by
\begin{align}
\xi_0^o\circ\bd{C}\ =&\
\sum_{n,l=0}^\infty\sum_{m,j,k=0}^\infty
\bigg(\id_m \otimes \Big(\id^{\otimes j}\otimes \xi_0^o\circ C_{n+1,l}\otimes \id^{\otimes k}\Big)\bigg)
\pi_{m+n+1,j+k+l},\\
\xi_0^o\circ C_{n+1,l}\ =&\ \frac{1}{l+1}\Bigg(
\xi_0^o C_{n+1,l}-(-1)^{\textrm{deg}(C)}\sum_{m=0}^{l-1}C_{n+1,l}\Big(\id_{n+1}\otimes
\big(\id^{\otimes(l-m-1)}\otimes\xi_0^o\otimes\id^{\otimes m}\big)\Big)
\Bigg)\,.
\end{align}
Then, Eq.~(\ref{app ocha 1}) at $p=0$\,,
\begin{align}
 \bd{C}^{(1)}(s)\ =&\ [\bd{Q},\bd{\nu}^{(1)}(s)]
\nonumber\\
&\
+ [\left(\bd{A}^{(0)}(s)+\bd{C}^{(0)}(s)\right),\bd{\nu}^{(1)}(s)]_{\mathfrak{o}_1(s)}
+ [\bd{C}^{(0)}(s),\bd{\mu}^{(1)}(s)]_{\mathfrak{o}_1(s)}
\nonumber\\
&\
+ [\bd{B}^{(0,0)}(s,s),\bd{\nu}^{(1)}(s)]_{\mathfrak{c}_1^0(s,s)}
+ [\bd{C}^{(0)}(s),(\bd{\lambda}^{(1,0)}+\bar{\bd{\lambda}}^{(0,1)})(s,s)]_{\mathfrak{c}_1^0(s,s)}
\nonumber\\
&\
+ s [\bd{B}^{(0,0)}(s,s),\bd{C}^{(0)}(s)]^{22}_{\Xi+\bar{\Xi}}\,,
\label{c 1}
\end{align}
determines $\bd{C}^{(1)}(s)$\,. Next, we solve Eq.~(\ref{app ocha 2}) at $p=1$\,,
\begin{align}
 [\bd{\eta},\bd{\nu}^{(2)}(s)]\ =&\ \partial_s\bd{C}^{(1)}(s)
\nonumber\\
&\ 
+ [\left(\bd{A}^{(0)}(s)+\bd{C}^{(0)}(s)\right),\bd{\nu}^{(1)}]^2
+ [\bd{C}^{(0)}(s),\bd{\mu}^{(1)}]^2
\nonumber\\
&\
+ [\bd{B}^{(0,0)}(s,s),\bd{\nu}^{(1)}(s)]^{21+12}
+ [\bd{C}^{(0)}(s),(\bd{\lambda}^{(1,0)}+\bar{\bd{\lambda}}^{(0,1)})(s,s)]^{21+12}\,,
\label{eta p1}
\end{align}
as
\begin{align}
 \bd{\nu}^{(2)}(s)\ =\ \xi_0^o\circ\Big(
&\
\partial_s\bd{C}^{(1)}(s)
+ [\left(\bd{A}^{(0)}(s)+\bd{C}^{(0)}(s)\right),\bd{\nu}^{(1)}]^2
+ [\bd{C}^{(0)}(s),\bd{\mu}^{(1)}]^2
\nonumber\\
&\
+ [\bd{B}^{(0,0)}(s,s),\bd{\nu}^{(1)}(s)]^{21+12}
+ [\bd{C}^{(0)}(s),(\bd{\lambda}^{(1,0)}+\bar{\bd{\lambda}}^{(0,1)})(s,s)]^{21+12}
\Big)\,.\label{nu 2}
\end{align}
Equation (\ref{app ocha 1}) at $p=1$ determines $\bd{C}^{(2)}(s)$ as
\begin{align}
 2\bd{C}^{(2)}(s)\ =&\ [\bd{Q},\bd{\nu}^{(2)}(s)]
\nonumber\\
&\
+ [\left(\bd{A}^{(1)}(s)+\bd{C}^{(1)}(s)\right),\bd{\nu}^{(1)}]_{\mathfrak{o}_1(s)}
+ [\bd{C}^{(1)}(s),\bd{\mu}^{(1)}]_{\mathfrak{o}_1(s)}
\nonumber\\
&\
+ [\left(\bd{A}^{(0)}(s)+\bd{C}^{(0)}(s)\right),\bd{\nu}^{(2)}]_{\mathfrak{o}_1(s)}
+ [\bd{C}^{(0)}(s),\bd{\mu}^{(2)}]_{\mathfrak{o}_1(s)}
\nonumber\\
&\
+ [(\bd{B}^{(1,0)}+\bd{B}^{(0,1)})(s,s),\bd{\nu}^{(1)}(s)]_{\mathfrak{c}_1^0(s,s)}
+ [\bd{B}^{(0,0)}(s,s),\bd{\nu}^{(2)}(s)]_{\mathfrak{c}_1^0(s,s)}
\nonumber\\
&\
+ [\bd{C}^{(1)}(s),(\bd{\lambda}^{(1,0)}+\bar{\bd{\lambda}}^{(0,1)})(s,s)]_{\mathfrak{c}_1^0(s,s)}
\nonumber\\
&\
+ [\bd{C}^{(0)}(s,s),(\bd{\lambda}^{(2,0)}+\bd{\lambda}^{(1,1)}
+ \bar{\bd{\lambda}}^{(1,1)}+\bar{\bd{\lambda}}^{(0,2)})(s,s)]_{\mathfrak{c}_1^0(s,s)}
\nonumber\\
&\
+ s\, [\bd{B}^{(0,0)}(s,s),\bd{\nu}^{(1)}(s)]^{22}_{X+\bar{X}}
+ s\, [\bd{C}^{(0)}(s),(\bd{\lambda}^{(1,0)}+\bar{\bd{\lambda}}^{(0,1)})(s,s)]^{22}_{X+\bar{X}}
\nonumber\\
&\
+ [\bd{B}^{(0,0)}(s,s),\bd{C}^{(1)}(s)]^{22}_{\Xi+\bar{\Xi}}
+ [(\bd{B}^{(1,0)}+\bd{B}^{(0,1)})(s,s),\bd{C}^{(0)}(s)]^{22}_{\Xi+\bar{\Xi}}\,.
\label{c 2}
\end{align}
One can obtain any $\bd{C}^{(p)}(s)$ one wants
by repeating the procedure.

Finally, it makes sense to mention that if we specify the type and number of inputs, 
the procedure ends in finite steps. We can explicitly determine any $\bd{C}_{n+1,l}^{(p)}\mid^{2r}$
you want in order from the one with the smallest number of inputs\footnote{The closed string input
is counted as 2.}.
The one with the smallest number of inputs is the open-closed interaction:
\begin{align}
 \bd{C}_{1,0}^{(0)}(s)\ =&\ \bd{C}_{1,0}^{(0)}\mid^2 + s\,\bd{C}_{1,0}^{(0)}\mid^0\,,\\
\bd{C}_{1,0}^{(1)}(s)\ =&\ \bd{C}_{1,0}^{(1)}\mid^0\,,
\end{align}
with $\bd{\nu}_{1,0}^{(1)}(s)=\bd{\nu}_{1,0}^{(1)}\mid^0$\,.
They are determined by Eqs.~(\ref{nu 1}), and (\ref{c 1}) under the initial condition in Eq.~(\ref{ini OCHA}) as
\begin{align}
 \bd{C}_{1,0}^{(0)}\mid^2\ =&\ (\bd{N}_B)_{1,0}\mid^2\,,\qquad 
 \bd{C}_{1,0}^{(0)}\mid^0\ =\ (\bd{N}_B)_{1,0}\mid^0\,,\qquad 
\bd{C}_{1,0}^{(1)}\mid^0\ =\ X_0^o(\bd{N}_B)_{1,0}\mid^0\,,
\end{align} 
with $\bd{\nu}_{1,0}^{(1)}\mid^0=\xi_0^o\circ(\bd{N}_B)_{1,0}\mid^0$\,.
It is a little more non-trivial for $\bd{C}^{(p)}_{1,1}(s)$\,:
\begin{align}
 \bd{C}_{1,1}^{(0)}(s)\ =&\  \bd{C}_{1,1}^{(0)}\mid^4 + s\,\bd{C}_{1,1}^{(0)}\mid^2 + s^2\,\bd{C}_{1,1}^{(0)}\mid^0\,,
\\ 
 \bd{C}_{1,1}^{(1)}(s)\ =&\  \bd{C}_{1,1}^{(1)}\mid^2 + s\,\bd{C}_{1,1}^{(1)}\mid^0\,,\\ 
 \bd{C}_{1,1}^{(2)}(s)\ =&\ \bd{C}_{1,1}^{(2)}\mid^0\,,
\end{align}
with
\begin{align}
\bd{\nu}_{1,1}^{(1)}(s)\ =&\ \bd{\nu}_{1,1}^{(1)}\mid^2 + s\,\bd{\nu}_{1,1}^{(1)}\mid^0\,,\\
\bd{\nu}_{1,1}^{(2)}(s)\ =&\ \bd{\nu}_{1,1}^{(2)}\mid^0\,.
\end{align}
These are determined by Eqs.~(\ref{nu 1}), (\ref{c 1}), (\ref{nu 2}), and (\ref{c 2})
under the initial condition in Eq.~(\ref{ini OCHA}) as
\begin{align}
 \bd{C}_{1,1}^{(0)}\mid^4\ =&\ (\bd{N}_B)_{1,1}\mid^4\,,\quad
 \bd{C}_{1,1}^{(0)}\mid^2\ =\ (\bd{N}_B)_{1,1}\mid^2\,,\quad
 \bd{C}_{1,1}^{(0)}\mid^0\ =\ (\bd{N}_B)_{1,1}\mid^0\,,\\
 \bd{C}_{1,1}^{(1)}\mid^2\ =&\ [\bd{Q},\bd{\nu}^{(1)}_{1,1}\mid^2]
+ [(\bd{M}_B)_2\mid^2,\bd{\nu}_{1,0}^{(1)}\mid^0]
+ [(\bd{N}_B)_{1,0}\mid^2,\bd{\mu}_2^{(1)}\mid^0]\,,\\
 \bd{C}_{1,1}^{(1)}\mid^0\ =&\ [\bd{Q},\bd{\nu}^{(1)}_{1,1}\mid^0]
+ [(\bd{M}_B)_2\mid^0,\bd{\nu}_{1,0}^{(1)}\mid^0]
+ [(\bd{N}_B)_{1,0}\mid^0,\bd{\mu}_2^{(1)}\mid^0]\,,\\
 \bd{C}_{1,1}^{(2)}\mid^0\ =&\ \frac{1}{2}\left([\bd{Q},\bd{\nu}^{(2)}_{1,1}\mid^0]
+ [\bd{A}_2^{(1)}\mid^0,\bd{\nu}_{1,0}^{(1)}\mid^0]
+ [\bd{C}_{1,0}^{(1)}\mid^0,\bd{\mu}_2^{(1)}\mid^0]\right) \,,
\end{align}
with
\begin{align}
\bd{\nu}_{1,1}^{(1)}\mid^2\ =&\ \xi_0^o\circ(\bd{N}_B)_{1,1}\mid^2\,,\qquad
\bd{\nu}_{1,1}^{(1)}\mid^0\ =\ 2\,\xi_0^o\circ(\bd{N}_B)_{1,1}\mid^0\,,\\
\bd{\nu}_{1,1}^{(2)}\mid^0\ =&\ \xi_0^o\circ\bd{C}_{1,1}^{(1)}\mid^0\,.
\end{align}
Those for $\bd{C}^{(p)}_{2,0}(s)$ are similarly obtained
by Eqs.~(\ref{app ocha 2}), and (\ref{app ocha 1}) at $p=0,1,2$
under the initial condition in Eq.~(\ref{ini OCHA}). 
We can continue these steps as long as we need.

\sectiono{Composite string fields in open superstring field theory
}\label{App A}

In this Appendix we show that the pure-gauge string field $G_o(V)$ 
for the open superstring field theory with general $A_\infty$ structure 
is obtained in a similar way given in the heterotic string field theory
\cite{Berkovits:2004xh}.
The pure-gauge string field $G_o(V_o)$ is associated with a finite form of
the ``gauge transformation'' 
\begin{align}
 \delta_{\delta V_o} \Psi\ =&\ \pi_1\bd{L}^\eta_o\left(\frac{1}{1-\Psi}\otimes
\delta V_o\otimes\frac{1}{1-\Psi}\right)
\nonumber\\
=&\ \eta \Lambda_o + L^{\eta}_2(\Psi, \delta V_o) + L^{\eta}_2(\delta V_o, \Psi)
+ \cdots,
\end{align}
with the infinitesimal parameter $\delta V_o$, and is obtained by integrating 
along a straight line connecting $0$ and $V_o$ that we parameterize
as $\tau V_o$ with $0\le \tau \le 1$. Considering that the difference between 
$G_o(\tau V_o + d\tau V_o)$ and $G_o(\tau V_o)$ is an infinitesimal gauge transformation,
we obtain a differential equation
\begin{equation}
 \partial_{\tau}G_o(\tau V_o)\ =\ 
 \pi_1\bd{L}^\eta_o\left(\frac{1}{1-g G_o(\tau V_o)}\otimes V_o\otimes\frac{1}{1-g G_o(\tau V_o)}\right),
\end{equation}
where we introduced a coupling constant $g$ for convenience.The pure-gauge string field $G_o(V_o)$ 
corresponds to $G_o(\tau V_o)$ at $\tau=1$
1and is obtained by solving this differential equation with the initial condition $G_o(0)=0$.
Expanding $G_o$ in the power of $g$ as $G_o=\sum_{n=0}^\infty g^nG^{(n)}_o$, we can sequentially solve the equation.
The equation at $\mathcal{O}(g^0)$ is given by
$\partial_\tau G_o^{(0)}=\eta V_o$ and is integrated as $G_o~{(0)}=\tau\eta V_o$.
At $\mathcal{O}(g)$, the equation becomes 
\begin{equation}
 \partial_\tau G_o^{(1)}\ =\ L^\eta_2(\tau\eta V_o, V_o) + L^\eta_2(V_o, \tau\eta V_o)
\end{equation}
and is solved as
\begin{equation}
G_o^{(1)}\ =\ \frac{\tau^2}{2}\Big(L^\eta_2(\eta V_o, V_o) + L^\eta_2(V_o, \eta V_o)\Big).
\end{equation}

Similarly, we can find $G_o$ up to any order of $g$ we want:
\begin{align}
 G_o(V_o)\ =&\ \eta V_o 
+ \frac{1}{2}\Big(L^\eta_2(\eta V_o, V_o) + L^\eta_2(V_o, \eta V_o)\Big)
\nonumber\\
&\
+ \frac{1}{3}\Big(
L^\eta_3(\eta V_o,\eta V_o, V_o)
+ L^\eta_3(\eta V_o,V_o, \eta V_o)
+ L^\eta_3(V_o, \eta V_o,\eta V_o)\Big)
\nonumber\\
&\
+ \frac{1}{3!}\Big(
L^\eta_2(L^\eta_2(\eta V_o, V_o), V_o) + L^\eta_2(V_o, \eta V_o,V_o)
\nonumber\\
&\hspace{4cm}
+ L^\eta_2(V_o, L^\eta_2(\eta V_o, V_o)) + L^\eta_2(V_o, L^\eta_2(V_o, \eta V_o))
\Big) + \cdots.
\end{align}

In order to find an explicit form of associated string field $B_d(V_o)$
($d=\partial_t$, $\delta$ or $Q$), we consider
\begin{equation}
 \mathcal{I}(\tau)\ =\ 
\pi_1^o\bd{L}^\eta_o\left(
\frac{1}{1-G_o(\tau V_o)}\otimes B_d(\tau;V_o,dV_o)\otimes\frac{1}{1-G_o(\tau V_o)}\right)
- (-1)^d dG_o(\tau V_o),
\end{equation}
and its $\tau$ derivative
\begin{align}
 \partial_\tau\mathcal{I}(\tau)\ =&\ 
\pi_1^o\bd{L}^\eta_o\bigg(
\frac{1}{1-G(\tau V_o)}\otimes
\Big(
\partial_\tau B_d(\tau;V_o,dV_o)-\mathcal{J}(\tau)
\Big)\otimes
\frac{1}{1-G(\tau V_o)}
\bigg)
\nonumber\\
&\
- \pi_1^o\bd{L}^\eta_o\bigg(
\frac{1}{1-G(\tau V_o)}\otimes\mathcal{I}(\tau)\otimes
\frac{1}{1-G_o(\tau V_o)}\otimes V_o\otimes\frac{1}{1-G_o(\tau V_o)}
\nonumber\\
&\hspace{20mm}
+ (-1)^d\frac{1}{1-G(\tau V_o)}\otimes V_o\otimes
\frac{1}{1-G_o(\tau V_o)}\otimes\mathcal{I}(\tau)\otimes\frac{1}{1-G_o(\tau V_o)}
\bigg),
\label{ass diff}
\end{align}
where
\begin{align}
\mathcal{J}(\tau)\ =&\
dV_o
+\pi_1^o\bd{L}^\eta_o
\Bigg(\frac{1}{1-G(\tau V_o)}\otimes V_o\otimes
\frac{1}{1-G(\tau V_o)}\otimes B_d(\tau;V_o,dV_o)\otimes\frac{1}{1-G(\tau V_o)}
\nonumber\\
&\hspace{10mm}
-(-1)^d\frac{1}{1-G(\tau V_o)}\otimes B_d(\tau;V_o,dV_o)\otimes
\frac{1}{1-G(\tau V_o)}\otimes V_o\otimes\frac{1}{1-G(\tau V_o)}\Bigg). 
\end{align}
If $B_d(\tau;V_o,dV_o)$ satisfies the differential equation
\begin{equation}
\partial_\tau B_d(\tau;V_o,dV_o)=\mathcal{J}(\tau),
\label{associated}
\end{equation}
with the initial condition $B_d(0;V_o,dV_o)=0$, then
$\partial_\tau\mathcal{I}(\tau)$ is proportional to $\mathcal{I}(\tau)$ with $\mathcal{I}(0)=0$,
and thus $\mathcal{I}(\tau)=0$ for ${}^\forall t$ due to Eq.~(\ref{ass diff}). 
Since $\mathcal{I}(1)=0$ is nothing but the relation 
in Eq.~(\ref{dG}) characterizing the associated string field, we can obtain the associated field 
$B_d(V_o,dV_o)$ by solving the differential equation in Eq.~(\ref{associated}).
Expanding $B_d=\sum_{n=0}^\infty g^nB_d^{(n)}$ with scaling $G_o\rightarrow gG_o$, we find that
\begin{align}
 B_d(V_o,dV_o)\ =&\  dV_o 
+ \frac{1}{2}\Big(
L_2^\eta(V_o,dV_o)-L_2^\eta(dV_o,V_o)\Big)
\nonumber\\
&\
+ \frac{1}{3}\Big(
L_3^\eta(\eta V_o, V_o, dV_o+)+L_3^\eta(V_o, \eta V_o, dV_o)+L_3^\eta(\eta V_o, V_o, dV_o)
\nonumber\\
&\
-L_3^\eta(\eta V_o, dV_o, V_o+-)L_3^\eta(dV_o, \eta V_o, V_o)-L_3^\eta(\eta V_o, dV_o, V_o)\Big)
\nonumber\\
&\
+\frac{1}{3!}\Big(
L_2^\eta(V_o,L_2^\eta(V_o,dV_o)) - L_2^\eta(V_o,L_2^\eta(dV_o,V_o))
\nonumber\\
&\
-L_2^\eta(L_2^\eta(V_o,dV_o),V_o) + L_2^\eta(L_2^\eta(dV_o,V_o),V_o)
\Big) + \cdots.
\end{align}

\end{document}